\documentclass[a4,aps,amssymb,showpacs,11pt,eqsecnum,nofootinbib]{revtex4}

\usepackage{graphicx,graphics}
\usepackage{enumerate}
\usepackage{subfigure}
\usepackage{amsmath}
\usepackage{amsfonts}
\usepackage{amssymb}
\usepackage{amsthm}
\usepackage{psfrag}

\catcode`\@=11
\def\numberbysection{\@addtoreset{equation}{section}
         \def\theequation{\arabic{section}.\arabic{equation}}}
\numberbysection

\def\a{\alpha}

\def\b{\beta}

\def\R{{\cal R}}

\def\Z{\mathbb{Z}}
\def\la{\langle}
\def\ra{\rangle}

\def\be{\begin{equation}}
\def\ee{\end{equation}}
\def\bea{\begin{eqnarray}}
\def\eea{\end{eqnarray}}
\newcommand\egal{&\!\!=\!\!&}

\begin{document}

\title{Logarithmic two-point correlators in the Abelian sandpile model}

\author{V.S. Poghosyan$^{1}$, S.Y. Grigorev$^2$,  V.B. Priezzhev$^2$ and P. Ruelle$^1$}
\affiliation{
$^1$Institut de Physique Th\'{e}orique, Universit\'{e} catholique de Louvain,
B-1348 Louvain-La-Neuve, Belgium\\
$^2$Bogoliubov Laboratory of Theoretical Physics,
Joint Institute for Nuclear Research, 141980 Dubna, Russia
}
\begin{abstract}
We present the detailed calculations of the asymptotics of two-site correlation functions for
height variables in the two-dimensional Abelian sandpile model. By using combinatorial methods for the enumeration of spanning trees, we extend the well-known result for the correlation $\sigma_{1,1} \simeq 1/r^4$ of minimal heights $h_1=h_2=1$ to $\sigma_{1,h} = P_{1,h}-P_1P_h$ for height values $h=2,3,4$. These results confirm the dominant logarithmic behaviour $\sigma_{1,h} \simeq (c_h\log r + d_h)/r^4 + {\cal O}(r^{-5})$ for large $r$, predicted by logarithmic conformal field theory based on field identifications obtained previously. We obtain, from our lattice calculations, the explicit values for the coefficients $c_h$ and $d_h$ (the latter are new).

\end{abstract}

\maketitle


\section{Introduction}

Conformal field theory (CFT) has proved to be a powerful tool in the
description of universality classes of equilibrium critical models
in two dimensions \cite{cft}. It has successfully helped to
understand and compute universal quantities and behaviours such as
critical exponents, correlation functions, finite-size scaling,
perturbations around fixed points and boundary conditions, among
others. More recently, increased interest has been turned toward a
larger class of conformal theories, namely the logarithmic conformal
field theories (LCFT), so called because they involve scaling fields
with inhomogeneous scaling transformations, which contain
logarithmic terms. They are believed to describe the continuum limit
of certain non-equilibrium lattice models, like dense polymers
\cite{polym,log}, sandpile models \cite{sand,diss,bound,jpr}, dimer
models \cite{dim} and percolation \cite{perc,log}, as well as the
infinite series of the lattice models recently defined in
\cite{log}. The distinctive characteristic of these lattice models
is the fact that they all have intrinsic non-local features; these
are thought to be responsible for the appearance, in the continuum
local theory, of logarithms in correlation functions. These
non-local features also mean that exact calculations on the lattice
are notoriously hard, thereby making a direct comparison with
logarithmic conformal theory predictions equally hard.

In this regard, we believe that the two-dimensional Abelian sandpile model is one of the lattice models
where the full consequences of the logarithmic conformal invariance can be most thoroughly and transparently tested and exploited.
All checks that could be carried out have been successful so far (see \cite{sand,diss,bound,jpr});
these include various correlators in the bulk and on boundaries, the effects of boundary conditions,
the determination boundary condition changing fields, the insertion of isolated dissipation and some finite-size effects.

Up to now, the only known sources of logarithms in correlators are the insertion of dissipation at isolated sites
\cite{diss} and the height variables for $h \geq 2$ \cite{jpr}.
The former is somewhat trivial but nonetheless in full agreement with the principles of LCFT
(in addition, the introduction of dissipation is in certain instances absolutely crucial as will be seen in Section \ref{sec4});
the logarithms produced in the presence of dissipation are not due to non-local features of the lattice model
but follow quite simply from the fact that the inverse Laplacian in two dimensions behaves logarithmically at large distances.
In contrast, joint probabilities with heights $h \geq 2$ are intrinsically non-local, and therefore much more important to check.

Despite the mentioned triviality of the logarithmic nature of the inverse Laplacian or Green function $G=\Delta^{-1}$, its non-local properties can be viewed by means of a geometrical construction proposed in \cite{IvPriez}. It was shown in \cite{IvPriez} that $G_{i,j}$ is related to the
number of two-rooted (and therefore two-component) spanning trees, such that both points $i,j$ belong to the same one-rooted subtree. The size $R$ of the one-component subtree imbedded into the spanning tree is distributed as $P(R) \sim 1/R$. Therefore, the Green function $G_{i,j}$ for two points separated by a distance $r$ is proportional to the integral over all components of size $R$ exceeding $r$. Below, we will see that this property of the Green function is relevant for the appearance of logarithmic corrections in joint height probabilities.

The only probabilities with heights $h \geq 2$ that have been computed so far are the 1-site probabilities on the upper-half plane.
Because 1-site height probabilities on the upper-half plane may be viewed as chiral 2-site joint probabilities on the full plane,
these could be used to make definite predictions as to the field theoretic nature of the bulk height variables in the scaling limit.
In this way, it was predicted in \cite{jpr} that the height variable $h=1$ goes, in the scaling limit,
to a primary field $\phi$ with conformal weights $(1,1)$, while the higher height variables $h=2,3,4$ scale to a unique field $\psi$
(up to normalization) which is the logarithmic partner\footnote{This is true only for the bulk height variables.
The boundary height variables on an open or closed boundary do not
scale to logarithmic fields, and consequently have no logarithm in their correlators \cite{bound}.}
of $\phi$ in a specific LCFT with central charge $c=-2$.
Moreover, this logarithmic theory turns out to be distinct from the well-known (and best understood) symplectic fermion (triplet) theory \cite{symp}.

If correct, these statements imply that the height 1 variable is the only one to have purely algebraic correlators.
Any multisite correlator involving at least one height variable $h \geq 2$ necessarily contains logarithmic functions of the separation distances,
powers of logarithms in case more than one height $h \geq 2$ is involved. Whereas this has been explicitely proved for general $n$-point,
bulk correlators of height 1 variables \cite{sand}, no comparable check has been done when a height $h \geq 2$ is present.

In this article, our purpose is to take a first step in this direction. We present here the detailed calculations of 2-site joint probabilities announced in \cite{PhysLetterB}.
Specifically, we compute, on the infinite discrete plane, the 2-site correlation functions $P_{1,h}(r)-P_1P_h$ of two height variables, one of which is $1$, the other is $h=2, 3$ or $4$, with $P_h$ the 1-site probability of height $h$ on the infinite plane. Given the field identifications conjectured in \cite{jpr}, conformal field theory predicts that the dominant term of the correlation functions is
\begin{equation}
\sigma_{1,h}(r) \equiv P_{1,h}(r) - P_1 P_h = {c_h \log r + d_h \over r^4} +  \ldots\,, \qquad h \geq 2,
\end{equation}
for some coefficients $c_h$ and $d_h$ \cite{jpr}. The explicit lattice calculations, to be detailed below, fully confirm these result, and exactly establish the values of coefficients $c_h$ and $d_h$.

The same arguments can be used to infer that the 2-site correlations of two heights bigger or equal to 2 decay like ${\log^2 r/r^4}$, but due to the complexity of the required combinatorics, the explicit lattice calculation of these correlations remains out of range for the moment.


\section{The Abelian sandpile model}
\label{sec2}

We consider the Abelian Sandpile Model (ASM) on a two-dimensional square grid $\mathcal{L}$.
A random variable $h_i$, which takes the integer values $1, 2, 3, 4, ...$, is attached to each site $i$,
representing the height of sand at that site.
A configuration $\mathcal C$ of the sandpile at a given time is the set of values $\{h_i,\, i\in \mathcal{L}\}$ for all sites.
The system starts its evolution from some initial state.
In this paper we are going to investigate observable quantities in the bulk, when the thermodynamic limit is taken and
the dependence on boundary details vanishes.
So we can start with square lattice with open boundary conditions for convenience.

The discrete time dynamics with open boundary conditions is completely defined in terms of the toppling matrix $\Delta$,
chosen to be the discrete Laplacian on $\mathcal L$:
\begin{equation}
\Delta_{ij} = 
\begin{cases}
\vspace{-3mm}
  4 & \text{if $i=j$},\\
\vspace{-3mm}
 -1 & \text{if $i,j$ are nearest neighbours}, \\
  0 & \text{otherwise}.
\end{cases}
\end{equation}

A configuration is called stable if all height values satisfy $h_i \leq 4$ for all $i \in \mathcal{L}$. At time $t$, the dynamics transforms the stable configuration ${\mathcal C}_t$ into a new stable configuration ${\mathcal C}_{t+1}$ as follows.
We add a grain of sand at a random site $i \in \mathcal{L}$ by setting $h_i \to h_i + 1$
(we assume that the site $i$ is chosen randomly with a uniform distribution on the grid $\mathcal L$).
This new configuration, if stable, defines ${\mathcal C}_{t+1}$.
If $h_i$ is bigger than 4, the site $i$ becomes non-stable and loses four grains of sand,
while all neighbours of $i$ receive one grain.
Note that if the site $i$ is on an (open) boundary where the number of neighbours is less than 4,
the system loses a corresponding number of grains.
Such sites are called dissipative.
In terms of the toppling matrix, if the site $i$ topples, the heights change according to
\begin{equation}
h_j \to h_j - \Delta_{ij}, \qquad \forall j \in {\mathcal L}.
\label{update}
\end{equation}
After site $i$ has toppled, other sites may become unstable, in which case they topple too,
according to the same toppling rule (\ref{update}).
Once all unstable sites have been toppled, a new stable configuration ${\mathcal C}_{t+1}$ is obtained.
One can show \cite{dhar-prl} that this dynamics is well-defined:
the order in which the unstable sites are toppled does not matter, and the new stable configuration ${\mathcal C}_{t+1}$ is reached after a finite number of topplings, provided some of the sites (at least one) are dissipative.

The long time behavior of the sandpile is described by the time invariant probability measure.
It assigns all stable configurations their probability of occurrence,
when the dynamics has been applied for a sufficiently long time so that the system has set in the stationary regime.

As such, the invariant measure one reaches might depend on the initial distribution.
In the present case however, there is no dependence on the initial distribution because there is a unique invariant measure,
$P_{\mathcal L}$ \cite{dhar-prl}.
The thermodynamic limit of $P_{\mathcal L}$ is what we want to compare with a conformal field theoretic measure.

The number of stable configurations is $4^{|\mathcal L|}$, but only a small fraction of them keep reappearing under the dynamical evolution.
The transient configurations exist at the initial stage of evolution and occur only a finite number of times. As a consequence,
they all have a zero measure with respect to $P_{\mathcal L}$.
The non-transient configurations are called recurrent and asymptotically occur with a non-zero probability.
Dhar has shown that the recurrent configurations all occur with equal probability under the ASM dynamics,
and that their total number is $\mathcal{N}=\det\Delta$ \cite{dhar-prl}.

A practical way to test whether a configuration is recurrent or transient is to use the ``burning'' algorithm \cite{dhar-prl}.
Given a recurrent configuration, this algorithm outputs the path followed by the fire to burn the configuration,
which in turn defines a unique rooted spanning tree on ${\mathcal L}$ \cite{majdhar2}.


\section{Height probabilities}
\label{sec3}

It has been shown in \cite{Priez} (see also \cite{jpr} for details) that the problem of computing the 1-site height probabilities
$P_h$, $h=1,2,3,4$ can be reduced to the problem of enumerating certain classes of spanning trees on $\mathcal L$.
We assume that all branches of the spanning tree are directed to its root, located somewhere on the boundary.
Formally, the determinant of the Laplacian $\Delta$ enumerates spanning forests (many-component trees) with roots on open boundary sites.
These spanning forests can be considered as spanning trees, if we add an auxiliary site and connect all boundary sites with it.
In the thermodynamic limit this boundary effect vanishes for the bulk.

We will say that a site $i_0$ is reachable from site $i$ if the unique directed path on the spanning tree from $i$ to the root passes through $i_0$.
In this case $i$ is called a predecessor of $i_0$.
Then the height probabilities at site $i_0$ are given by
\begin{equation}
P_1 =       \frac{X_0}{4\, \mathcal{N}};\quad
P_2 = P_1 + \frac{X_1}{3\, \mathcal{N}};\quad
P_3 = P_2 + \frac{X_2}{2\, \mathcal{N}};\quad
P_4 = P_3 + \frac{X_3}{    \mathcal{N}}.
\label{siteprob}
\end{equation}
where $X_k$ with $k=0,1,2,3$ is the number of spanning trees on $\mathcal L$ such as the site
$i_0$ has exactly $k$ nearest neighbour predecessors \cite{Priez}.

\begin{figure}[b]
\includegraphics[width=30mm]{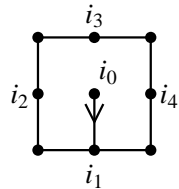}
\vspace{-1.2cm}
\caption{\label{fig1} $\Delta'$ is related to $\Delta$ by removing the bonds connecting $i_0$
to three of its neighbours. The remaining bond connecting $i_0$ to its neighbourhood
can be oriented toward $i_1$ (above), $i_2,\, i_3$ or $i_4$.
}
\end{figure}

Let us consider the probability $P_1$. The quantity $X_0$ is the number of spanning trees in which the reference site $i_0$
is connected to only one of its neighbours.
This means that $i_0$ is reachable from no other site in $\mathcal L$, or equivalently, $i_0$ is a leaf.
The simplest way to compute $X_0$ is to remove the bonds connecting $i_0$ to three of its neighbours,
for instance $i_2,i_3,i_4$ as shown in Fig.\ \ref{fig1}, and then apply Kirchhoff's theorem to a new toppling matrix $\Delta'$.
One may write $\Delta' = \Delta + B_1$, where the matrix $B_1$ has non-zero elements only in rows and columns labelled by $i_0,i_2,i_3,i_4$:
\begin{eqnarray}
\nonumber
& & \hspace{5mm}
\begin{array}{cccc}
\; i_0 & \;\; i_2  & \;\; i_3  & \;\; i_4\\
\end{array}\\
B_1&=&
\left(
\begin{array}{rrrr}
-3 & 1  & 1  &  1\\
 1 & -1 & 0  &  0\\
 1 & 0  & -1 &  0\\
 1 & 0  & 0  & -1\\
\end{array}
\right)
\begin{array}{c}
i_0 \\ i_2 \\ i_3 \\ i_4
\end{array}      \label{SandPile-B1}
\end{eqnarray}
%
%

One then obtains $X_0=4\det\Delta'$, since the remaining bond can take four different orientations, and finally
\begin{equation}
P_1 = \frac{\det\Delta'}{\det\Delta} = \det(I + B_1G),
\label{P1_det}
\end{equation}
where the matrix $G = \Delta^{-1}$ is the inverse of the toppling matrix. Since the grid $\mathcal L$ is finite,
the matrices $B_1, \Delta$ and $G$ are finite too; $P_1$ is given by a finite determinant and depends explicitly on the location of $i_0$.

In the thermodynamic limit, i.e. in the limit of infinite lattice $\left({\mathcal L} \to \Z^2\right)$,
$G$ goes to the inverse Laplacian or Green function on the full (discrete) plane.
The size of the matrix $B$ becomes infinite but retains a rank equal to 4, which implies that the determinant in (\ref{P1_det})
reduces to a $4\times4$ non-trivial determinant.
The explicit form of the translationally invariant Green function on the plane is
\begin{equation}
G_{\vec{r}_1,\vec{r}_2} \equiv G(\vec{r}_2-\vec{r}_1) \equiv G_{0,0} + g_{p,q}, \quad \vec{r}_2-\vec{r}_1\equiv\vec{r}\equiv(p,q)
\end{equation}
with $G_{0,0}$ an irrelevant infinite constant. The (finite) numbers $g_{p,q}$ are given explicitly by
\begin{equation}
g_{p,q}= \int\!\!\!\!\int_{-\pi}^{\pi} \frac{{\rm d} \alpha {\rm d} \beta}{8\pi^2} \;
\frac{e^{ {\rm i}\, p \, \alpha + {\rm i} \, q \, \beta}-1}{2-\cos\alpha-\cos\beta}.
\label{Green}
\end{equation}
Let us mention symmetry properties of this function:
\begin{equation}
g_{p,q}=g_{q,p}=g_{-p,q}=g_{p,-q}.
\label{GreenSymmetry}
\end{equation}
After the integration over $\alpha$, it can be expressed in a more convenient form for actual calculations,
\begin{equation}
g_{p,q} = \frac{1}{4\pi} \int_{-\pi}^{\pi} {\rm d} \beta \; \frac{t^p \, e^{ {\rm i}\, q\, \beta} - 1 }{ \sqrt{y^2-1} },
\label{Green2}
\end{equation}
where $t = y - \sqrt{y^2-1}$, $y = 2 - \cos{\beta}$. For $r^2 = p^2 +q^2 \gg 1$ it has a behaviour \cite{3-leg}
\begin{equation}
\label{GreenFunctionAsymptAppendix}
g_{p,q} = -\frac{1}{2\pi}\left(\log r + \gamma +\frac{3}{2}\log 2\right)
+\frac{\cos (4\, \varphi )}{24\, \pi\,  r^2}
+\frac{18 \cos (4\, \varphi )+ 25 \cos (8\, \varphi )}{480\, \pi\, r^4} + \ldots,
\end{equation}
where $(p,q) = (r\cos \varphi,r\sin \varphi)$, and with $\gamma = 0.57721...$ the Euler constant. The values of $g_{p,q}$ for certain small $p$ and $q$, needed for the calculations below, can be found in \cite{spitz}.

In the thermodynamic limit, the $4\times4$ determinant in (\ref{P1_det}) no longer depends on a position of $i_0$ by translational invariance,
and yields \cite{majdhar1}
\begin{equation}
P_1=\frac{2(\pi-2)}{\pi^3} \simeq 0.07363.
\label{P1}
\end{equation}

\begin{figure}[t]
\includegraphics[width=135mm]{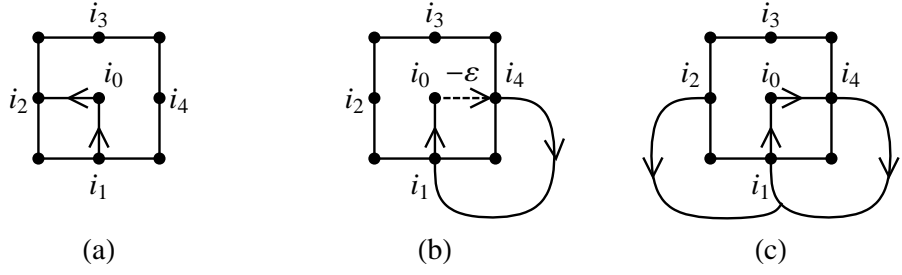}
\caption{\label{fig2} Pictorial decomposition of $X_1$ in terms of local, loop and theta configurations.
(a) The diagram of the spanning trees constrained to have an arrow from $i_1$ to $i_0$, and from $i_0$ to $i_2$,
and no arrow between $i_0$ and $i_3, i_4$.
(b) The diagram of the arrow configurations which have a single loop containing arrows form $i_1$ to $i_0$, from $i_0$ to $i_4$,
and no arrow from $i_2,\, i_3$ to $i_0$.
The diagram (c) corresponds to the subclass of configurations in (b) which have a path of arrows starting from $i_2$ and ending at $i_1$.}
\end{figure}

While the calculation of $X_0$ is relatively easy, that of $X_1,X_2$ and $X_3$ is much harder.
The basic reason for this is that the spanning trees involved in $X_0$ are subjected to a local constraint,
while those counting for $X_k$, $k>0$, must satisfy a global constraint. In the case of $X_1$, for instance,
a spanning tree is counted if the reference site is reachable from one and only one nearest neighbour.
To illustrate the problem, consider the derivation of $P_2$, which requires $X_1$ only.
According to \cite{Priez}, $X_1$ is expressed by three terms,
\begin{equation}
\label{X1}
X_1 = 3 \sum_D \left(\mathcal{N}_{\rm local}^D-\mathcal{N}_{\rm loop}^D + \mathcal{N}_{\Theta}^D \right).
\end{equation}
Each term corresponds to a particular class of arrow configurations, as shown in Fig.\ \ref{fig2}.
The index $D=N,W,S,E$ denotes four possible global orientations of the diagrams in Fig.\ \ref{fig2}.
The coefficient 3 accounts for the possible directions of the outgoing arrow at $i_0$ once the incoming arrow is fixed. In a situation  where the rotational symmetry holds, like the one considered here, the numbers $\mathcal{N}_{\rm local}^D,\, \mathcal{N}_{\rm loop}^D,\, \mathcal{N}_{\Theta}^D$ do not depend on $D$ so we can fix the incoming arrow at $i_0$ (say from $S$ direction, like in Fig.\ \ref{fig2}).
Then, from (\ref{siteprob}) and (\ref{X1}) it yields
\begin{equation}
\label{N1}
P_2 =P_1 + 4\left.\left(
\frac{\mathcal{N}_{\rm local}^D}{\mathcal{N}}-
\frac{\mathcal{N}_{\rm loop}^D}{\mathcal{N}} +
\frac{\mathcal{N}_{\Theta}^D}{\mathcal{N}}\right)\right|_{D=S}.
\end{equation}

The term $\mathcal{N}_{\rm local}^D$ for the orientation $D=S$ is the number of spanning trees,
in which the site $i_0$ has only one incoming arrow and it is an arrow from the site $i_1$, the outgoing arrow from $i_0$ being directed to $i_2$.
Since these are purely local conditions, the number of such spanning trees can be calculated in the way used for $X_0$. To do this, we introduce a defect matrix $B_{\rm local}$ whose effect is to remove the bonds $(i_0,i_3)$, $(i_0,i_4)$ and the possible outgoing arrows from $i_1$ to its three nearest neighbours $i_5 = i_1 - \hat e_x$, $i_6 = i_1 + \hat e_x$ and $i_7 = i_1 - \hat e_y$.
The non-zero elements of $B_{\rm local}$ can be collected in a finite matrix:
\begin{eqnarray}
\nonumber
& & \hspace{6mm}
\begin{array}{ccccccc}
i_0 & \;\; i_3 & \;\; i_4 & \;\; i_1 & \;\;\; i_5 & \;\;\; i_6 & \;\;\; i_7
\end{array}\\
B_{\rm local} & = &
\left (
\begin{array}{rrrrrrr}
-3 &  1 & 1  & 1  & \quad 0 & \quad 0 & \quad 0 \\
 1 & -1 & 0  & 0  & \quad 0 & \quad 0 & \quad 0 \\
 1 &  0 & -1 & 0  & \quad 0 & \quad 0 & \quad 0 \\
 0 &  0 & 0  & -3 & \quad 1 & \quad 1 & \quad 1
\end{array} \right)
\begin{array}{c}
i_0 \\ i_3 \\ i_4 \\ i_1
\end{array},
\end{eqnarray}

Using the expression similar to (\ref{P1_det}), we obtain
\begin{equation}
\label{Nlocal}
\frac{\mathcal{N}_{\rm local}^D}{\mathcal{N}} = \det(I + B_{\rm local}G) = \frac{1}{2\pi}-\frac{5}{2\pi^2}+\frac{4}{\pi^3}.
\end{equation}

The term $\mathcal{N}^D_{\rm loop}$ is the non-local first contribution.
It counts all those configurations in the first term for which the site $i_4$ is a predecessor of $i_0$
due to a path from $i_4$ to $i_1$ and an arrow from $i_1$ to $i_0$.
To find it, we introduce an extra arrow $i_0$ to $i_4$ with weight $-\varepsilon$ and remove incoming arrows from
$i_2$, $i_3$, $i_4$ to $i_0$.
In the limit $\varepsilon \to +\infty$ (see \cite{Priez}) the determinant of the so-constructed defect matrix does not enumerate spanning trees
but the arrow configurations having a single loop (with weight $-1$) containing the oriented bonds
$(i_1,i_0)$ and $(i_0,i_4)$, and no arrow from $i_2$, $i_3$ and $i_4$ to $i_0$, as shown in Fig.\ \ref{fig2}(b).
The extra arrow with weight $-\varepsilon$ on the bond $(i_0,i_4)$ is shown by the dashed line.
We multiply the determinant by $-1$ to cancel the unwanted weight $-1$ of the loop:
\begin{eqnarray}
\nonumber
& & \hspace{3mm}
\begin{array}{cccc}
i_0 &\;\; i_4 &\;\; i_3 &\;\; i_2
\end{array}\\
\frac{ \mathcal{N}^D_{\rm loop} }{ \mathcal{N} } = - \lim_{\varepsilon \to \infty} \frac{1}{\varepsilon} \det (I + B_{\rm loop}G),
\quad \hbox{with }
B_{\rm loop} & = &
\left (
\begin{array}{rrrr}
0 & -\varepsilon &  0 &  0 \\
1 &      -1      &  0 &  0 \\
1 &       0      & -1 &  0 \\
1 &       0      &  0 & -1
\end{array} \right)
\begin{array}{c}
i_0 \\ i_4 \\ i_3 \\ i_2
\end{array}.
\end{eqnarray}
Evaluating the finite determinant one yields
\begin{equation}
\label{Nloop}
\frac{\mathcal{N}_{\rm loop}^D}{\mathcal{N}}=-\frac{1}{4\pi^2} + \frac{2(\pi-2)}{\pi^3} \, G_{0,0}.
\end{equation}
It contains a term proportional to the infinite constant $G_{0,0}$
(on a finite lattice, $G_{0,0}$ diverges as $\log{L}$ with the size $L$ of the system).
This divergence in (\ref{Nloop}) reflects the fact that the number of diagrams with
loops is much greater than the number of spanning trees, in a proportion that diverges in the infinite volume limit.
Since $X_1$ is a finite fraction of the total number of spanning trees,
the factor $G_{0,0}$ present in $\mathcal{N}_{\rm loop}^D/\mathcal{N}$
must be canceled by a similar term in the third term $\mathcal{N}^D_{\Theta}/\mathcal{N}$.

The subtraction effected by the second term removes too many configurations, namely those for which the path from $i_2$ does not go to the root, but goes to the loop.
The aim of the third term is to restore these configurations.

\begin{figure}[b]
\includegraphics[width=100mm]{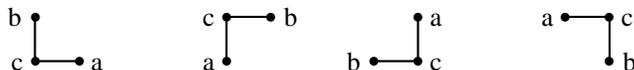}
\vspace{-7mm}
\caption{\label{fig-abc} Four possibilities for the junction $a,b,c$.}
\end{figure}

The calculation of $\mathcal{N}^D_{\Theta}$ for the $\Theta$-graphs is more complicated and truly non-local.
For the configurations shown in Fig.\ \ref{fig2}(c), the two legs starting from $i_2$ and $i_4$ come together at a certain site $c$.
Around $c$, the junction takes one of the following shapes, $\perp$, $\vdash$, $\top$ or $\dashv$\,,
with two incoming arrows and one outgoing arrows at $c$ (the central site in Fig. \ref{fig-abc}).
Since for each orientation of the pattern $\perp$, there are three ways to place the three arrows,
this makes a total of twelve different local configurations around the site $c$.

\begin{figure}[b]
\includegraphics[width=130mm]{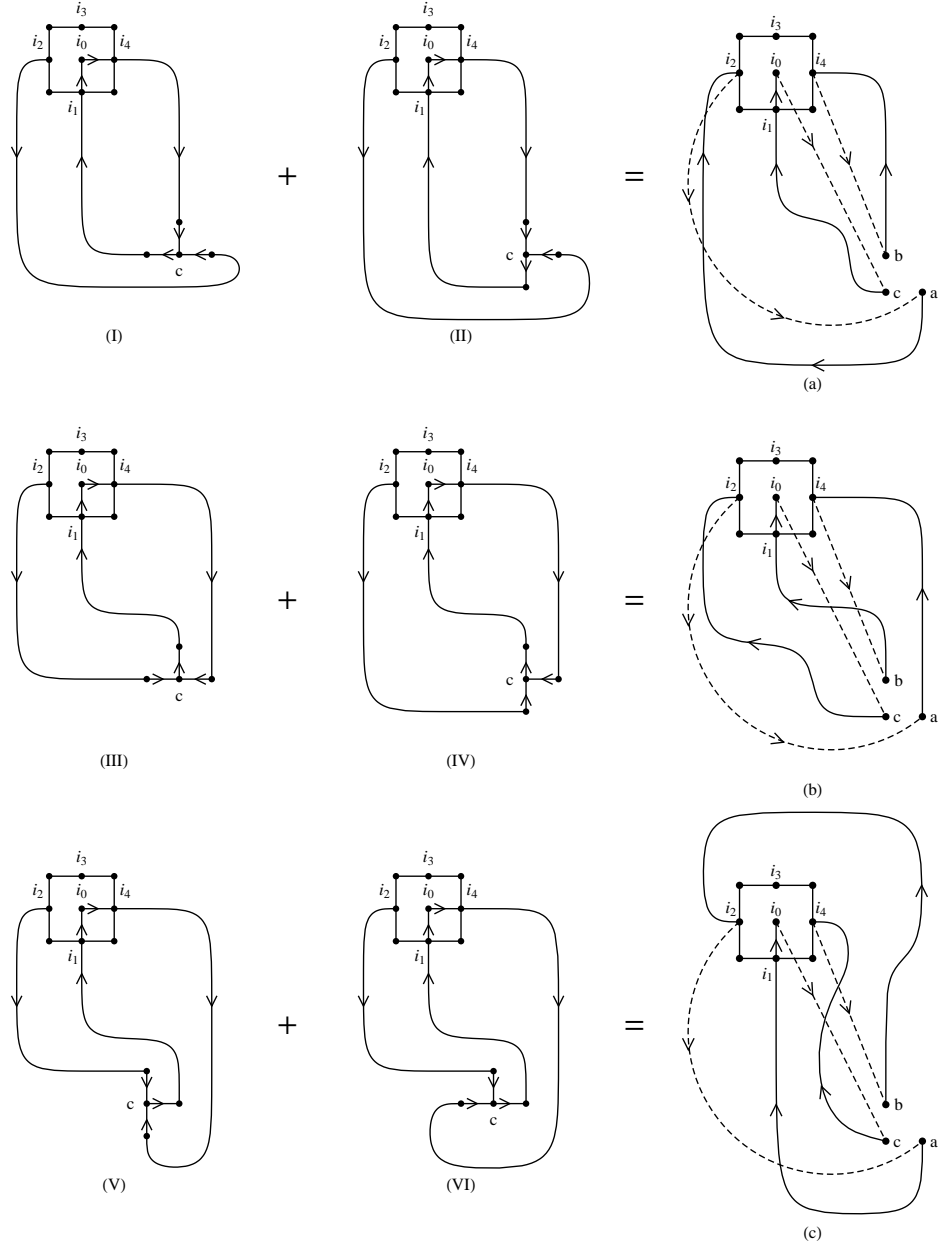}
\vspace{-6mm}
\caption{\label{figTL} In (i) to (vi), schematic representation of six of the twelve $\Theta$-graphs that need be computed in order to evaluate $\mathcal{N}_\Theta$.
They reduce to the three diagrams on the right, marked (a) to (c), where the dashed lines represent the inserted bridges;
all three bridge connections must be part of a loop.}
\end{figure}

Six of them, namely those with the basic pattern oriented as $\perp$ and $\vdash$\,, are shown in Fig.\ \ref{figTL}(i-vi).
The six others are associated to the patterns $\dashv$ and $\top$.
To compute the contributions of these twelve classes of diagrams, the so-called ``bridge'' trick has been devised \cite{Priez},
by which some of the arrows in the original diagram are moved so as to make bridges between sites around $i_0$ and sites around $c$.
We give here the details of the procedure for the six diagrams in Fig.\ \ref{figTL}(i-vi), the other six being similar.

We start with the two classes of $\Theta$-graphs shown in Fig.\ \ref{figTL}(i,ii).
If, in both diagrams, we remove the three arrows $(i_0 \to i_4$), ($a \to c$) and ($b \to c$),
and add new long ranged links (bridges) $(i_0 \to c$), ($a \to i_2$) and ($b \to i_4$), we obtain a link structure with three separate loops.
If we further change the orientations of the two loops containing the sites $a$ and $b$, and this is a totally harmless change,
we obtain the link diagram shown in Fig.\ \ref{figTL}(a). Note that the two diagrams in Fig.\ \ref{figTL}(i,ii),
which differ by the direction of the arrow coming out from $c$, are both counted in Fig.\ \ref{figTL}(a),
since there the direction of this arrow has not been specified (and can only be to the west or to south).
The correspondence goes in both ways, so that we actually have, refering to the diagrams in Fig.\ \ref{figTL}, that (a) = (i) $+$ (ii).
The configurations with the structure (a) are defined on a modified lattice since we added three new connections between the sites
$i_0,i_2,i_4$ and $c,a,b$ respectively.
Once these three connections are introduced, the configurations with the structure (a) are characterized by the presence of three
loops with each of the three new connections contained in a separate loop (at this stage, the positions of $a,b,c$ are fixed;
they will summed over later). We proceed similarly for the two remaining pairs of diagrams, although their topology yields slightly different results.

For the diagrams in Fig.\ \ref{figTL}(iii,iv), we remove the arrows ($a \to c$) and ($c \to b$) and add the links ($a \to i_2$) and ($c \to i_0$),
thereby creating a loop ($i_0 \to a \to i_2 \to c \to i_0$).
Reversing the orientation of this loop, removing the arrow ($i_4 \to i_0$) and adding the link ($i_4 \to b$)
results in the link structure in Fig.\ \ref{figTL}(b).
Contrary to the previous case, the so-obtained structure contains a single loop ($i_0 \to c \to i_2 \to a \to i_4 \to b \to i_0$).
Thus the arrow configurations in Fig.\ \ref{figTL}(b) use the same new connections as Fig.\ \ref{figTL}(a),
and are characterized by the presence of the single loop just mentioned.

For the diagrams in Fig.\ \ref{figTL}(v,vi), we perform the following sequence of changes: we change the orientation of the original loop
($i_0 \to i_4 \to c \to i_0$); remove the arrow ($i_4 \to i_0$), add ($i_4 \to b$) and change the orientation of the new loop;
we remove the arrows ($a \to c$) and ($c \to b$), add ($a \to i_2$) and ($c \to i_0$), thus making a single loop containing $a,b,c$;
finally we change the orientation of this loop to obtain the structure shown in Fig.\ \ref{figTL}(c).
It is characterized by the presence of a single loop ($i_0 \to c \to i_4 \to b \to i_2 \to a \to i_0$).

In this way, we have reduced six diagrams to three diagrams, one containing three loops, the other two containing a single loop.
A further and crucial simplification is provided by the following observation: the union of the configurations
shown in Fig.\ \ref{figTL}(a-c), for fixed sites $a,b,c$, can be uniquely characterized by (i)
the absence of the arrow ($i_3 \to i_0$), since it was forbidden in the diagram Fig.\ \ref{fig2}(c), and (ii) the fact that all three arrows ($i_0 \to c$), ($i_2 \to a$) and ($i_4 \to b$) living on the three new connections must be contained in loops.
The second condition holds because the topology of the connections implies that the three arrows are either contained in three separate loops,
like in Fig.\ \ref{figTL}(a), or in a single loop, which then must be of the type shown in Fig.\ \ref{figTL}(b) or Fig.\ \ref{figTL}(c).

For fixed $a,b,c$ forming a corner (see Fig.\ \ref{fig-abc}) the defect matrix is
\begin{eqnarray}
\nonumber
& & \hspace{3mm}
\begin{array}{rrrrr}
i_0 &\;\; i_3 &\;\; a &\;\;\; b &\;\;\; c
\end{array}\\
B_\Theta & = &
\left(
\begin{array}{rrrrr}
 1 & -1&       0      &        0      &      0     \\
 0 & 0 & -\varepsilon &        0      &      0     \\
 0 & 0 &       0      &  -\varepsilon &      0     \\
 0 & 0 &       0      &        0      &-\varepsilon\\
\end{array}
\right)
\begin{array}{c}
i_3 \\ i_2 \\ i_4 \\ i_0
\end{array}.
\end{eqnarray}

The total number of $\Theta$-graphs is then obtained by summing over all
possible positions of the sites $[a,b,c]$ in the plane and all four orientations $\mathcal{D}$ of the ``head'' of the $\Theta$-graph around $i_0$,
\begin{equation}
\frac{\mathcal{N}_\Theta}{\mathcal{N}} =
-\frac{1}{2}\sum_{[a,b,c]}\sum_{\mathcal{D}} \lim_{\varepsilon \to \infty} \frac{1}{\varepsilon^3} \det \bigl( I +B_{\Theta} G\bigr),
\end{equation}
where the factor $1/2$ appears, since four orientations of points $a,b,c$ around $c$
take into account four junction possibilities $\perp$, $\vdash$, $\top$, $\dashv$ twice.
The summation over $[a,b,c]$ can be decomposed into double sum over all positions $(k,l)$ of the site $c$ and
a sum of all four orientations $\mathcal{D}_\llcorner=\llcorner,\ulcorner,\urcorner,\lrcorner$ of $[a,b,c]$ around $c$. Regarding the summation we have to keep in mind that the situations when points $a$, $b$ or $c$ overlap with $j_0$, $j_2$, $j_3$ or $j_4$ are forbidden and should be subtracted:
\begin{equation}
\label{Ntheta2}
\frac{\mathcal{N}_{\Theta}}{\mathcal{N}} = -\frac{1}{2}
\sum_{(k,l)}\sum_{\mathcal{D}_\llcorner}\sum_{\mathcal{D}}\lim_{\varepsilon \to \infty} \frac{1}{\varepsilon^3} \det \bigl( I +B_{\Theta} G\bigr) - F_{\Theta},
\end{equation}
where the correction term $F_\Theta$ excludes the few situations,
when the triplet of sites $[a,b,c]$ overlaps the head of the $\Theta$-graph (the vicinity of $i_0$). In \cite{Priez} it has been found to be
\begin{equation}
\label{Ftheta}
F_{\Theta} = -\frac{1}{8} + \frac{7}{8\pi} - \frac{5}{4\pi^2} - \frac{2(\pi - 2)}{\pi^3} \,G_{0,0}.
\end{equation}
{}From (\ref{Nloop}) and (\ref{Ftheta}), we see that the terms
proportional to $G_{0,0}$  cancel. The infinite double summation over
$(k,l)$ in (\ref{Ntheta2}) can be carried out explicitly, but the
final result remains expressed as a double integral of a complicated function
\cite{Priez}. A simple and exact formula for $P_2$ has however been
conjectured \cite{jpr}, based on a high precision numerical evaluation of the double integral,
\begin{equation}
\label{P2_final}
P_2=\frac{1}{4}-\frac{1}{2\pi}-\frac{3}{\pi^2}+\frac{12}{\pi^3} \simeq 0.1739.
\end{equation}

\begin{figure}[t]
\includegraphics[width=100mm]{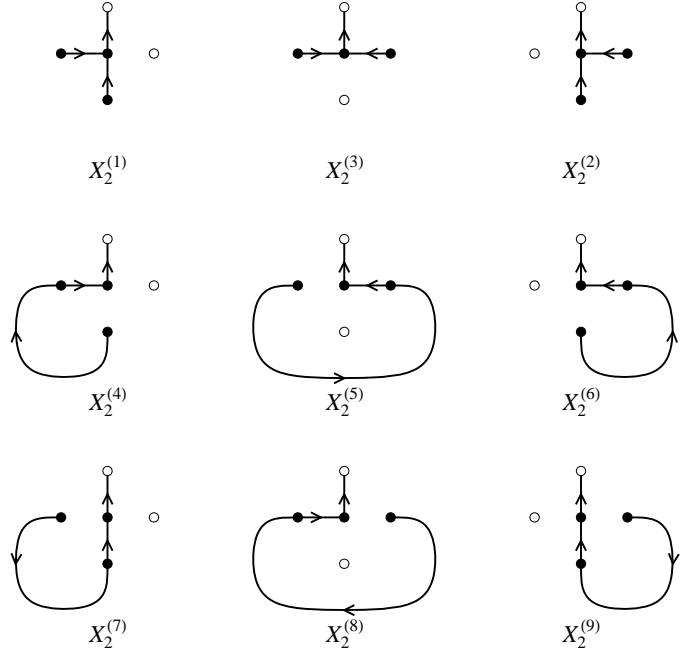}
\vspace{-7mm}
\caption{\label{fig11} All diagrams $X_2$ for fixed arrow
from $i_0$ to $i_4$. A black site means that this site is a predecessor of $i_0$ (there
is a path of arrows from this site to the root through $i_0$), and a white site indicates that the site is not a predecessor of $i_0$. }
\end{figure}

\begin{figure}[h!]
\includegraphics[width=135mm]{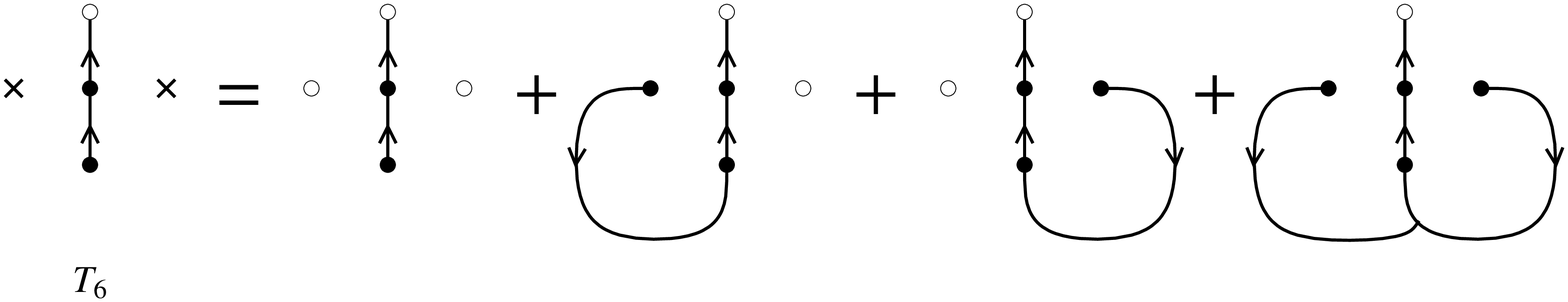}
\includegraphics[width=120mm]{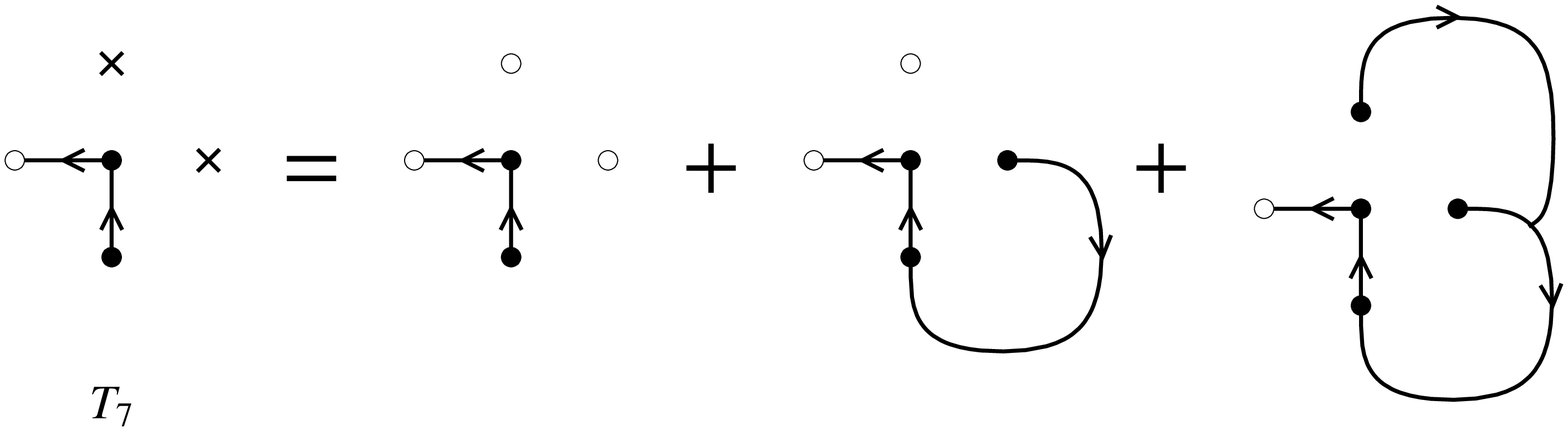}
\vspace{-5mm}
\caption{\label{fig12And13} Graphical representation of (\ref{X24X26X27X29ToT7T6}).
Sites marked by a cross may be or may not be predecessor of $i_0$, however they cannot be direct predecessors of $i_0$ (their outgoing arrow pointing directly to $i_0$).}
\end{figure}

\begin{figure}[t]
\includegraphics[width=135mm]{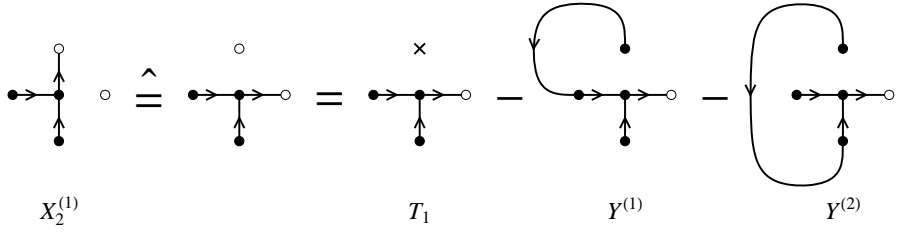}
\vspace{-5mm}
\caption{\label{X21AndX23ToTYY} Diagrams of type $X_2^{(1)}$.}
\end{figure}

\begin{figure}[t]
\includegraphics[width=120mm]{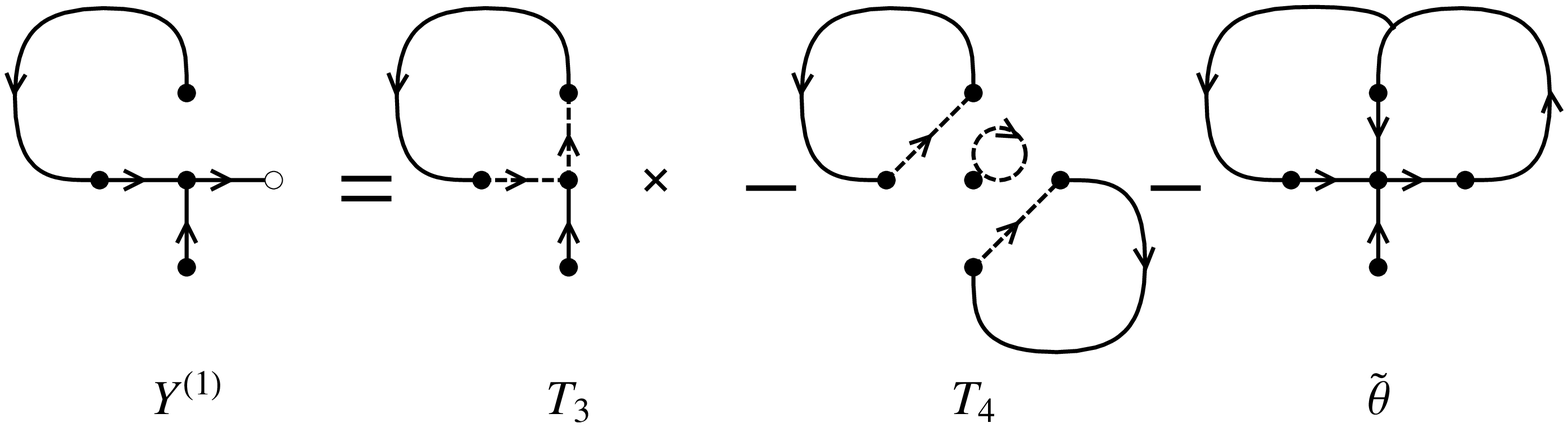}
\vspace{-5mm}
\caption{\label{YY1} Diagrams of type $Y^{(1)}$.}
\end{figure}

\begin{figure}[t!]
\includegraphics[width=110mm]{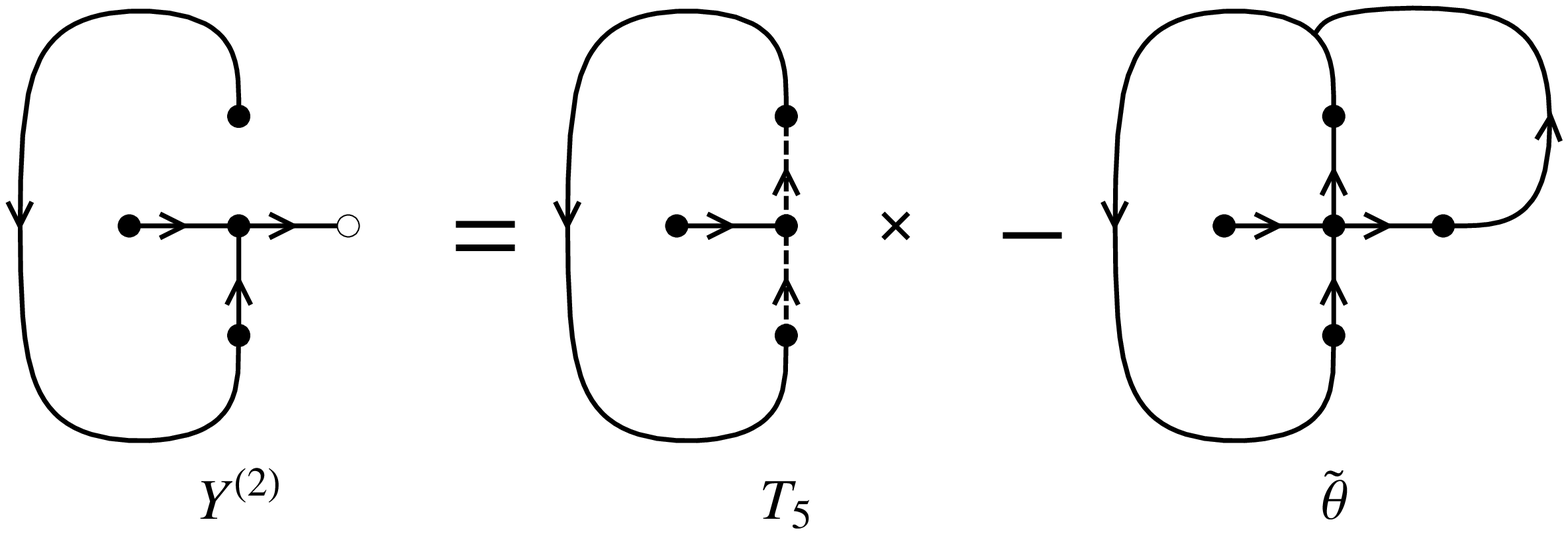}
\vspace{-6mm}
\caption{\label{YY2} Diagrams of type $Y^{(2)}$.}
\end{figure}

\begin{figure}[t!]
\includegraphics[width=110mm]{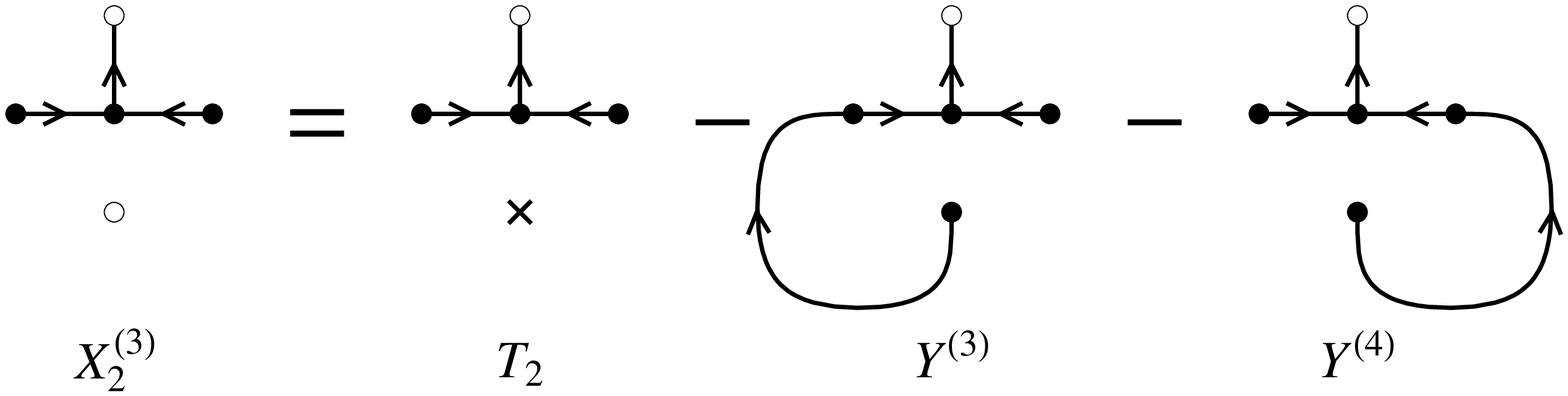}
\vspace{-6mm}
\caption{\label{X22ToYY} Diagrams of type $X_2^{(3)}$.}
\end{figure}

The height 3 probability $P_3$ is expressed by diagrams for $X_2$ shown in Fig.\ \ref{fig11}. Diagrams of type $X_2^{(5)}$ and $X_2^{(8)}$ are impossible for topological reasons. Four of the diagrams can be calculated as,
\begin{equation}
\label{X24X26X27X29ToT7T6}
X_2^{(4)}+X_2^{(6)}+X_2^{(7)}+X_2^{(9)} \; \hat{=} \; 4(T_6-T_7),
\end{equation}
where $T_6$ and $T_7$ are defined in Fig.\ \ref{fig12And13}.
In fact they are diagrams of the same type but with different
orientations $D$ or reflections $R$, for example $X_2^{(4)}$ and
$X_2^{(6)}$. They may take different values in
the presence of a fixed defect, nevertheless after summation over
all orientations $D$ and reflections $R$, we obtain an equality, 
\begin{equation}
\label{sumDRToHatEqual}
\sum_{D,R}X_2^{(4)}(D,R)=\sum_{D,R}X_2^{(6)}(D,R).
\end{equation}
For such configurations, we use the equality sign with a ``hat'', that is $X_2^{(4)}\:\hat{=}\:X_2^{(6)}\:\hat{=}\:X_2^{(7)}\:\hat{=}\:X_2^{(9)}$. We also have $X_2^{(1)}\:\hat{=}\: X_2^{(2)}$ and reduce them to the diagrams shown in Fig.\ \ref{X21AndX23ToTYY}. The diagrams $Y^{(1)}$ and $Y^{(2)}$ can be reduced to those shown in Fig.\ \ref{YY1} and \ref{YY2} correspondingly. Similarly the diagram $X_2^{(3)}$ is reduced to the diagrams in Fig.\ \ref{X22ToYY}, where $Y^{(3)}\:\hat{=}\: Y^{(4)}\:\hat{=}\: Y^{(1)}$.

\begin{figure}[t]
\includegraphics[width=40mm]{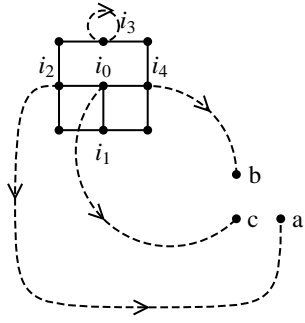}
\vspace{-2mm}
\caption{\label{fig9} The schematic pattern of the $\widetilde{\Theta}$-graph.}
\end{figure}

\begin{figure}[t!]
\includegraphics[width=130mm]{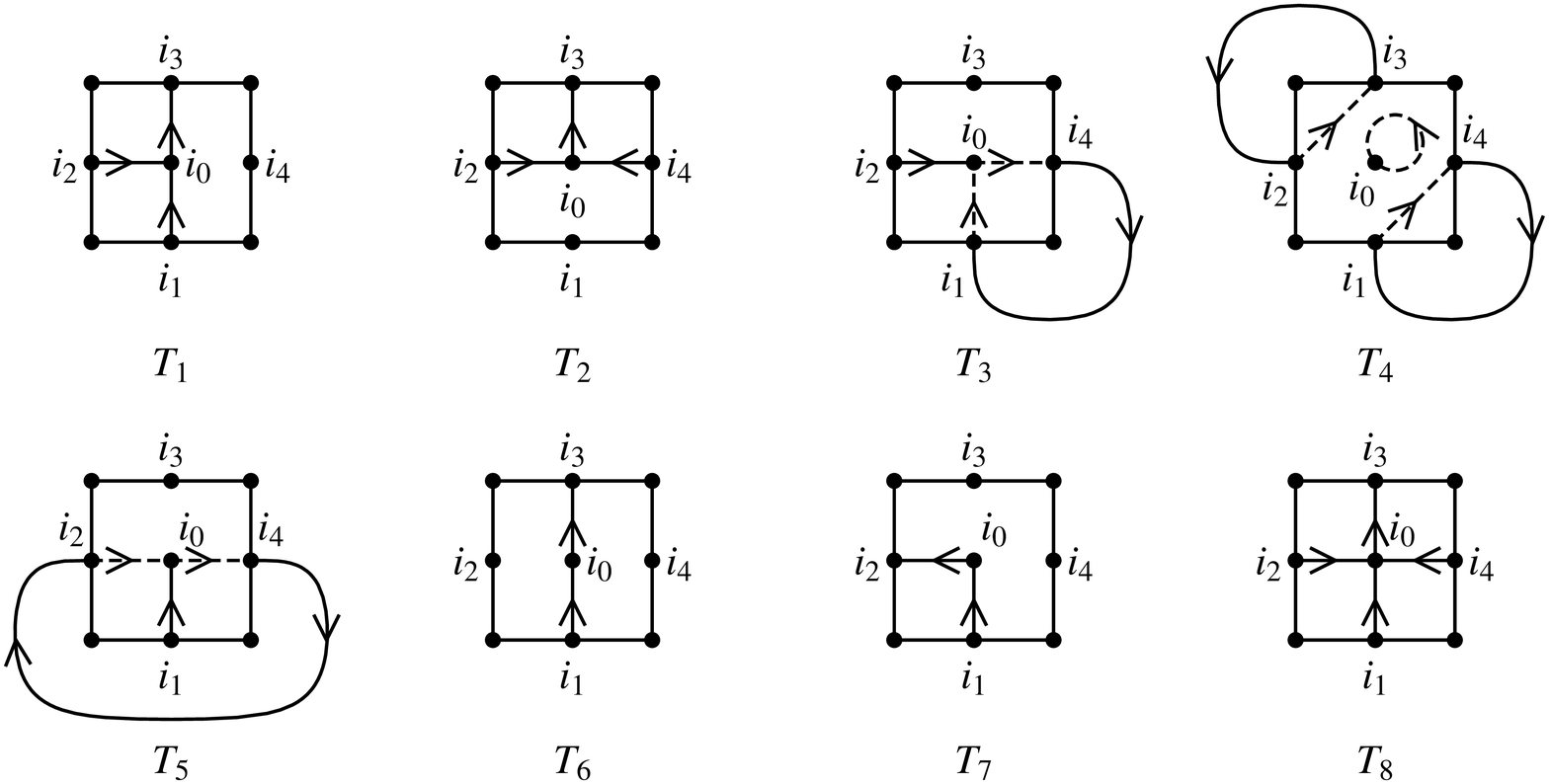}
\vspace{-5mm}
\caption{\label{Tconfig} Diagrams needed for calculating $P_3$ and $P_4$.}
\end{figure}

Eventually, $P_3$ is expressed in terms of the diagrams shown in Fig.\
\ref{fig9} and Fig.\ \ref{Tconfig}
\begin{equation}
\label{P3_NandT}
P_3=P_2+\frac{1}{2} \left[6\frac{\mathcal{N}_{\widetilde{\Theta}}}{\mathcal{N}}+
2\frac{T_1}{\mathcal{N}}+\frac{T_2}{\mathcal{N}}-4\frac{T_3}{\mathcal{N}}+4\frac{T_4}{\mathcal{N}}-2\frac{T_5}{\mathcal{N}}
+4\frac{T_6}{\mathcal{N}}-4\frac{T_7}{\mathcal{N}}  \right],
\end{equation}
where we have introduced another type of the theta graph denoted by $\widetilde{\Theta}$ (Fig.\ \ref{fig9}) and a defect matrix
\begin{equation}
\label{BTheta} B_{\widetilde{\Theta}}= \left(
\begin{array}{ccccc}
 -\varepsilon&  0&  0&  0\\
            0& -\varepsilon&  0&  0\\
            0&  0& -\varepsilon&  0\\
            0&  0&  0& -\varepsilon\\
\end{array}
\right).
\end{equation}


We may compute the height 4 probability $P_4$ through similar diagrams and check the identity $\sum_{i=1}^4 P_i=1$. More simply, we can use the last identity for computing $P_4$.
Like for $P_2$, the final expressions for $P_3$ and $P_4$ also involve complicated double integrals \cite{Priez},
conjectured in \cite{jpr} to reduce to the following simple expressions:
\begin{eqnarray}
\label{P3andP4}
P_3&=&\frac{3}{8}+\frac{1}{\pi}-\frac{12}{\pi^3} \simeq 0.3063, \\
\label{P3andP4_2}
P_4&=&\frac{3}{8}-\frac{1}{2\pi}+\frac{1}{\pi^2}+\frac{4}{\pi^3} \simeq 0.4462.
%
\end{eqnarray}


\section{Predictions of conformal field theory for pair correlation functions}
\label{sec4}

During recent years, considerable evidence \cite{sand,diss,bound,jpr} has been gathered which strongly support the following assertions: (i) the Abelian sandpile model is conformally invariant in the scaling limit; (ii) conformal invariance is realized by a logarithmic conformal field theory,
where some of the local fields belong to  indecomposable Virasoro representations, and consequently satisfy logarithmic scaling \cite{lcft}; and (iii) the relevant logarithmic theory has central charge $c=-2$, and is distinct from the symplectic free fermion theory,
also known as the triplet theory \cite{symp}.
We briefly review here what has been found concerning the height variables and their description by scaling fields.

In the Abelian sandpile model, the height $h_z$ at a bulk site $z$ can take four values. We accordingly consider four random lattice variables,
\begin{equation}
h_a(z) = \delta(h_z-a)-P_a\,, \qquad a=1,2,3,4.
\label{delta}
\end{equation}
The quantities $P_a$ are the bulk one-site height probabilities in the thermodynamic limit, as computed in the previous section, and we normalize the variables $h_a(z)$ to have a zero expectation value in the infinite volume limit. In the scaling limit, the variables $h_a(z)$ converge to conformal fields $h_a(z,\bar z)$, whose nature has been determined in \cite{jpr}.

Not surprisingly, in view of the calculations of the previous section, the height 1 field is very different from the other height fields.
It was found that $h_1$ is a primary field with conformal weights $(1,1)$, while the other three, $h_2,h_3$ and $h_4$,
are all related to a single field, identified with the logarithmic partner of $h_1$.

More precisely, consider the triplet of fields $\psi, \phi, \rho$ of conformal weights $(1,1),(1,1),(0,1)$,
satisfying the following relations under chiral conformal transformations,
\begin{eqnarray}
L_0 \psi \egal \psi - \frac{1}{2} \phi, \quad L_1 \psi = \rho, \quad L_{-1} \rho = -\frac{1}{4} \phi, \quad (L_0 - 1) \phi = L_0 \rho = 0,\label{L0} \\
L_p \phi \egal L_p \rho = 0, \quad (p \geq 1), \qquad (L^2_{-1} - 2L_{-2}) \phi = 0,
\end{eqnarray}
and similar relations for antichiral conformal generators (they however involve a new field $\bar \rho$ with weights (1,0)).
The factor $-\frac{1}{2}$ in the first relation in (\ref{L0}) is conventional and fixes the relative normalizations of $\phi$ and $\psi$,
unlike the factor $-\frac{1}{4}$ in the third relation which is an intrinsic parameter, independent of the field normalizations.
The field $\phi$ is primary, and degenerate at level 2, whereas, in contrast, $\psi$ has an inhomogeneous transformation under dilations.
In fact, $\phi$ and $\psi$ are members of a non-chiral version of the indecomposable representation called $\R_{2,1}$ in \cite{gk1}.
The fact that $\psi$ has an inhomogeneous scaling transformation readily implies that correlators containing several $\psi$'s will contain logarithms.

The four height fields $h_a(z,\bar z)$ have been identified in terms of $\phi$ and $\psi$ in \cite{jpr},
based on the calculation of the one-point probabilities $P_a(z)$ on the upper-half plane. The results are as follows.
The height one and two fields, $h_1(z,\bar z)$ and $h_2(z,\bar z)$ are equal to $\phi$ and $\psi$ respectively,
while the other two are linear combinations of $\phi$ and $\psi$,
\begin{eqnarray}
h_1(z,\bar z) &\!=\!& \phi(z,\bar z)\,,\\
h_2(z,\bar z) &\!=\!& \psi(z,\bar z)\,,\\
h_3(z,\bar z) &\!=\!& \alpha_3 \psi(z,\bar z) + \beta_3 \phi(z,\bar z)\,,\\
h_4(z,\bar z) &\!=\!& \alpha_4 \psi(z,\bar z) + \beta_4 \phi(z,\bar z)\,.
\end{eqnarray}
The normalizations of these fields are those of the corresponding lattice random variables.
Specifically, the normalization of $\phi$ will be fixed from the lattice calculation of $P_{1,1}(r)$,
and in turn fixes that of $\psi$ from the above algebraic relations\footnote{The full normalization of $\psi$ involves two constants,
since if $\psi$ satisfies the algebraic relations, any combination $\alpha \psi + \beta \phi$ will satisfy them as well.
The normalization of $\phi$ fixes $\alpha$ but not $\beta$. This freedom is manifest in the correlators (\ref{phiphi})-(\ref{psipsi}).}.
The values of the coefficients $\alpha_3, \beta_3, \alpha_4,\beta_4$ are then chosen so that $h_3$ and $h_4$ reproduce the normalizations
of their lattice counterparts. Evidently the four fields are not independent, but satisfy $h_1 + h_2 + h_3 + h_4 = 0$.

The identification of the height fields makes it possible to compute correlations, and then compare them with lattice calculations.
The definition (\ref{delta}) of the variables $h_a(z)$ makes it obvious that
\begin{equation}
\la h_a(z) h_b(w) \ra_{\rm ASM} = P_{a,b}(z-w) - P_a \, P_b,
\end{equation}
and naturally suggests that the scaling limit of this lattice correlation corresponds to the 2-point function
$\la h_a(z,\bar z) h_b(w,\bar w) \ra$, computed in the conformal theory.
A convincing argument that this cannot be the case is to notice that $\la \phi(z,\bar z) \phi(w,\bar w)\ra = 0$,
an easy-to-prove and well-known fact in logarithmic conformal theories, would imply $P_{1,1}(z-w) = P_1^2$ in the scaling limit,
in contrast to what the explicit calculation yields \cite{majdhar1}.

If $P_{a,b}(z_{12})$ denotes the joint probability that the height at $z_1$ be $a$ and
the height at $z_2$ be $b$, the 2-site correlations are given by
$P_{a,b}(z_{12}) - P_aP_b$ for $a,b=1,2,3,4$. In the scaling regime, when the
distance $z_{12}$ is large, these correlations should be equal to expectation
values of pairs of fields $h_a(z_1,\bar z_1) h_b(z_2,\bar z_2)$. However the plane
has no boundary where sand can leave the system, so that the prescription we used
earlier requires to insert a bulk dissipation field $\omega(\infty)$ at infinity.
In fact the correct correspondence states that the scaling limit of the lattice correlation is in terms of a 3-point correlator,
\begin{equation}
{\rm scalim} \:[P_{a,b}(z-w) - P_a \, P_b] = \la h_a(z,\bar z) h_b(w,\bar w) \omega(\infty) \ra,
\end{equation}
where $\omega$ is a weight (0,0) conformal field, logarithmic partner of the identity \cite{jpr}.
Indeed we have stressed in Section \ref{sec2} that, for the dynamics of the model to be well-defined,
the finite lattice should include dissipative sites, and that they were all located on the boundary.
Since the infinite plane can be thought of as the limit of a sequence of growing finite grids, it means that,
in the infinite volume limit, the boundaries, and with them, the dissipation, are sent off to infinity.
The field $\omega(\infty)$ precisely realizes the insertion of dissipation at infinity \cite{diss}, required for the sandpile model to be well-defined.

The relevant 3-point correlators have been computed in \cite{jpr}, and take the form:
\begin{eqnarray}
\la \phi(z_1,\bar z_1) \phi(z_2,\bar z_2) \omega(\infty) \ra \egal \frac{A}{|z_{12}|^4}\,, \label{phiphi} \\
\la \phi(z_1,\bar z_1) \psi(z_2,\bar z_2) \omega(\infty) \ra \egal \frac{1}{|z_{12}|^4} \Big\{A \log{|z_{12}|} + B \Big\}\,,\\
\la \psi(z_1,\bar z_1) \psi(z_2,\bar z_2) \omega(\infty) \ra \egal \frac{1}{|z_{12}|^4} \Big\{A \log^2{|z_{12}|} + 2B \log{|z_{12}|} + C\Big\}\,,\label{psipsi}
\end{eqnarray}
with $z_{12} \equiv z_1-z_2$.

From these results and the above relation between the height fields and $\phi,\psi$, we easily obtain the required correlators,
$\la h_a(z_1,\bar z_1) h_b(z_2,\bar z_2) \omega(\infty) \ra$ for large $r$:
\begin{eqnarray}
\sigma_{a,b}(r) = P_{a,b}(r) - P_a P_b&\!=\!& \frac{1}{r^4} \Big\{\a_a\a_b C + (\a_a\b_b + \b_a\a_b) B + \b_a \b_b A \nonumber\\
&\!+\!&\; [2\a_a\a_b B + (\a_a\b_b+\b_a\a_b)A] \log{r} + \a_a\a_b A \log^2{r}\Big\} + \ldots\,.
\label{pijgen}
\end{eqnarray}

The LCFT arguments predict a relation between chiral two-point correlation functions and height probabilities in the presence of a boundary.
Their explicit asymptotic expressions at large distance $r$ from the boundary have been obtained in \cite{jpr}
and the values of coefficients $\alpha_a$ and $\beta_a$ computed. Two types of boundaries were considered -- open and closed.
Namely, it has been shown that the coefficients $\alpha_a$ and $\beta_b$ take the values
\begin{equation}
  \begin{array}{ll}
\alpha_1 =  0, &\quad \beta_1 = 1,\\
\alpha_2 =  1, &\quad \beta_2 = 0,\\
\alpha_3 =  \frac{8-\pi}{2(\pi-2)}, &\quad \beta_3 = -\frac{48 - 12\pi + 5\pi^2 - \pi^3}{4(\pi-2)^2},\\
\alpha_4 = -\frac{\pi+4}{2(\pi-2)}, &\quad \beta_4 =  \frac{32 +  4\pi +  \pi^2 - \pi^3}{4(\pi-2)^2}.
  \end{array}
\label{alpha}
\end{equation}
The value of the constant $A$ can be found from a known result by Majumdar and Dhar on minimal height two-point correlation function \cite{majdhar1}
\begin{equation}
\sigma_{1,1}(r) = -\frac{P_1^2}{2\, r^4} + \ldots\, , \quad r \gg 1,
\end{equation}
from which it follows that
\begin{equation}
A = -\frac{P_1^2}{2} = - \frac{2(\pi-2)^2}{\pi^6}.
\label{Constant-A}
\end{equation}
One cannot find the other constants $B$ and $C$ within conformal field theory.
Numerical simulations suggest the values $B = -0.0045 \pm 0.0005$ and $C = -0.009 \pm 0.0005$ \cite{jpr}.

In the rest of this paper, we will proceed to a check of the correlations (\ref{pijgen})
by means of the explicit calculation of $\sigma_{1,a}$, $a=2,3,4$, whose form should be
\begin{equation}
\sigma_{1,a}(r) = \frac{\alpha_a B + \beta_a A}{r^4} + \frac{\alpha_a A \log{r}}{r^4} + \ldots\, .
\label{sigma1a}
\end{equation}
This form $\sigma_{1,a}(r)$, as well as the coefficients $\alpha_a$, $\beta_a$ and $A$ will be checked from lattice calculations,
and verified to be in agreement with the values given above. We will compute the exact value of $B$, to find it equal to
\begin{equation}
B = -\frac{P_1^2}{2} \left(\gamma + \frac{3}{2}\log{2}\right) -\frac{(\pi-2)(16-5\pi)}{\pi^6} = -0.0047305.
\label{Bconst}
\end{equation}
The constant $C$ appears in the subdominant term $1/r^4$ in the correlation functions $\sigma_{a,b}$ for heights $a,b=2,3,4,$
and is well out of reach for the moment.


\section{Lattice calculations of $\sigma_{1,2}(r)$}

\begin{figure}[t]
\includegraphics[width=70mm]{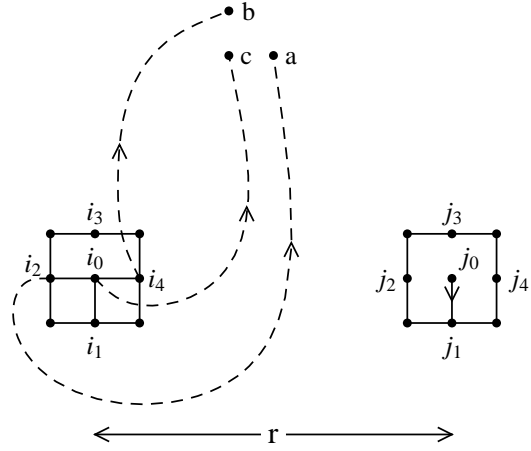}
\vspace{-3mm}
\caption{\label{fig4} Geometric set-up for the calculation of the correlation $P_1$ and a $\Theta$-graph.}
\end{figure}

\begin{figure}[t]
\includegraphics[width=120mm]{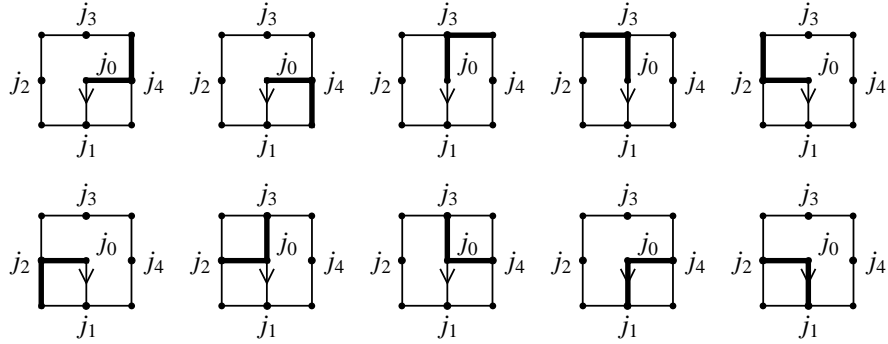}
\vspace{-3mm}
\caption{\label{fig6} Forbidden configurations, when the points $a,b,c$ (forming the thick-lined angle) overlap with the defect located at $j_0$.}
\label{fig-forbidden}
\end{figure}

Consider two sites $i_0$ and $j_0$ on the lattice, a distance $r$ apart.
The joint probability $P_{1,2}(r)$ to have a height $2$ at $i_0$ and height $1$ at $j_0$ can be found by the method analogous to that used in Section III.
Namely, we have
\begin{equation}
\label{P12-local-loop-theta}
P_{1,2}(r) = P_{1,1}(r) + \sum_D \left(
\frac{\mathcal{N}_{1, \rm local}^D(r)}{\mathcal{N}}-
\frac{\mathcal{N}_{1, \rm loop }^D(r)}{\mathcal{N}}+
\frac{\mathcal{N}_{1, \Theta   }^D(r)}{\mathcal{N}}
\right).
\end{equation}
The values $\mathcal{N}_{1, \rm local}^D(r)$, $\mathcal{N}_{1, \rm loop }^D(r)$ and $\mathcal{N}_{1, \Theta}^D(r)$ can be
calculated by defect matrices $B_{1,\rm local}$, $B_{1,\rm loop}$ and $B_{1,\Theta}$ if we combine the defect
$B_1$ around the site $j_0$ with the defects correspondingly $B_{\rm local}$, $B_{\rm loop}$ and $B_{\Theta}$ around $i_0$ (see Fig. \ref{fig4}). Calculating the determinants and using the expansion (\ref{GreenFunctionAsymptAppendix}), we obtain the following asymptotic expressions for large $r$,
\begin{eqnarray}
\sum_D \frac{\mathcal{N}_{1, \rm local}^D(r)}{\mathcal{N}} &\!\!=\!\!&
\frac{4 (\pi-2)(\pi^2-5\pi+8)}{\pi^6} - \frac{(\pi-2)(\pi-4)^2}{\pi^6 r^4} + \ldots\, ,\\
\sum_D \frac{\mathcal{N}_{1, \rm loop}^D(r)}{\mathcal{N}} &\!\!=\!\!&
\frac{16(\pi-2)^2}{\pi^6} G_{0,0}
-\frac{2(\pi-2)}{\pi^5} + \nonumber \\
&\!\!+\!\!&\frac{4(\pi-2)^2}{\pi^7 r^2}
-\frac{8(\pi-2)^2}{\pi^6 r^4} G_{0,0}
+\frac{(\pi-2)(\pi^2+4\pi-6)}{\pi^7 r^4}+\ldots\, .
\end{eqnarray}
The third term reads
\begin{equation}
\sum_D \frac{\mathcal{N}_{1, \Theta}^D(r)}{\mathcal{N}} = -\frac{1}{2}
\sum_{(k,l)}\sum_{\mathcal{D}_\llcorner}\sum_{\mathcal{D}}\lim_{\varepsilon \to \infty}
\frac{1}{\varepsilon^3} \det \bigl( I +B_{1,\Theta} G\bigr)
-F_{1,\Theta}^{i_0}-F_{1,\Theta}^{j_0} \, ,
\label{N1Theta}
\end{equation}
where $F_{1,\Theta}^{i_0}$ and $F_{1,\Theta}^{j_0}$ are expressions for the forbidden configurations,
when the triplet $[a,b,c]$ overlaps the vicinities of $i_0$ or $j_0$ correspondingly.
The first type of forbidden configurations are same as in the case of single-height probabilities \cite{Priez}. The forbidden configurations of second type $F_{1,\Theta}^{j_0}$ are schematically shown in Fig.\ref{fig-forbidden}. Direct calculations of all forbidden diagrams give
\begin{eqnarray}
F_{1,\Theta}^{i_0} &\!\!=\!\!&
-\frac{16 (\pi-2)^2}{\pi^6} G_{0,0}
-\frac{(\pi-5)(\pi-2)^2}{\pi^5}-\\
&\!\!-\!\!&\frac{4(\pi-2)^2}{\pi^7 r^2}
-\frac{8 (\pi-1) (\pi-2)^2}{\pi ^6 r^4} G_{0,0}
+\frac{(\pi-2)^2 (7\pi^2 + 2\pi - 12)}{4 \pi^7 r^4}+\ldots\,,
\end{eqnarray}
and
\begin{equation}
F_{1,\Theta}^{j_0} = \frac{4(\pi-2)^2}{\pi^6r^4} \left(2\pi G_{0,0} - 1 - \log r - \gamma - \frac{3}{2} \log 2 \right) + \ldots\,.
\end{equation}
The nonzero elements of the defect matrix $B_{1,\Theta}$ are
\begin{eqnarray}
\nonumber & & \hspace{0.2in}
\begin{array}{ccccccccc}
\hspace{-0.1in} i_0 &\;\; i_3 &\;\; c &\;\; a &\;\; b &\;\; j_0 &\;\; j_2 &\;\; j_3 &\;\; j_4 %
\end{array}\\
B_{1,\Theta}&=& \left(
\begin{array}{cccccccccc}
 1& -1& 0&  0&  0&  0&  0&  0&  0\\
 0&  0&-\varepsilon&  0&  0&  0&  0&  0&  0\\
 0&  0& 0& -\varepsilon&  0&  0&  0&  0&  0\\
 0&  0& 0&  0& -\varepsilon&  0&  0&  0&  0\\
 0&  0& 0&  0&  0& -3&  1&  1&  1\\
 0&  0& 0&  0&  0&  1& -1&  0&  0\\
 0&  0& 0&  0&  0&  1&  0& -1&  0\\
 0&  0& 0&  0&  0&  1&  0&  0& -1
 \end{array}
\right)
\begin{array}{c} i_3\\ i_0\\ i_2\\ i_4\\ j_0\\ j_2\\ j_3\\ j_4
\end{array}
\end{eqnarray}
and we have to compute the function
\begin{equation}
\sigma_{1,\Theta}(r) = -\frac{1}{2}
\sum_{(k,l)}\sum_{\mathcal{D}_\llcorner}\sum_{\mathcal{D}}\lim_{\varepsilon \to \infty} \left.\left.\frac{1}{\varepsilon^3}
\right(\det \left( I +B_{1,\Theta} G\right) - P_1 \det \left( I +B_{\Theta} G\right)\right).
\label{sigma1Theta}
\end{equation}
In section VI, we will show that $\sigma_{1,\Theta}(r)$ has the following expansion for large $r$,
\begin{equation}
\sigma_{1,\Theta}(r) = - \frac{3 P_1^2}{2 \, r^4}\left(\log r+\gamma+\frac{3}{2}\log 2\right)
+ \frac{(\pi-2)(11\pi^2-48\pi+80)}{4\pi^6 \, r^4} - \frac{(\pi-2)(16-5\pi)}{\pi^6 \, r^4} +\ldots\,,
\label{sigma1ThetaResult}
\end{equation}

From (\ref{P12-local-loop-theta}) and the expressions for $F_{1,\Theta}^{i_0}$ and $F_{1,\Theta}^{j_0}$, we have
\bea
\sigma_{1,2}(r) &=& \sigma_{1,\Theta}(r)
+ \frac{P_1^2}{r^4}\left(\log r+\gamma+\frac{3}{2}\log 2\right)
- \frac{(\pi-2)(11\pi^2-48\pi+80)}{4\pi^6 r^4} + \ldots\, \nonumber\\
&=& - \frac{P_1^2}{2 \, r^4}\left(\log r+\gamma+\frac{3}{2}\log 2\right) - \frac{(\pi-2)(16-5\pi)}{\pi^6 \, r^4} +\ldots\,,
\eea
which confirms the prediction (\ref{sigma1a}) for $a=2$, and yields the constant $B$ in (\ref{Bconst}).

The calculation of $P_{1,3}(r)$ involves the composition of
$\widetilde{\Theta}$-graph introduced in Section \ref{sec4} at site $i_0$ with the defect corresponding to $P_1$ at site $j_0$.
The nonzero elements of the corresponding defect matrix $B_{1,\widetilde{\Theta}}$ read

\begin{eqnarray}
\nonumber & & \hspace{0.3in}
\begin{array}{ccccccccc}
\hspace{-0.1in} i_3 &\;\; c &\;\; a &\;\; b &\;\;\; j_0 &\;\; j_2 &\;\; j_3 &\;\; j_4 %
\end{array}\\
B_{1,\widetilde{\Theta}}&=& \left(
\begin{array}{cccccccccc}
 -\varepsilon& 0&  0&  0&  0&  0&  0&  0\\
 0&-\varepsilon&  0&  0&  0&  0&  0&  0\\
 0& 0& -\varepsilon&  0&  0&  0&  0&  0\\
 0& 0&  0& -\varepsilon&  0&  0&  0&  0\\
 0& 0&  0&  0& -3&  1&  1&  1\\
 0& 0&  0&  0&  1& -1&  0&  0\\
 0& 0&  0&  0&  1&  0& -1&  0\\
 0& 0&  0&  0&  1&  0&  0& -1
 \end{array}
\right)
\begin{array}{c} i_3\\ i_0\\ i_2\\ i_4\\ j_0\\ j_2\\ j_3\\ j_4
\end{array}
\end{eqnarray}
The correlation
$\sigma_{1,\widetilde{\Theta}} = P_{1,\widetilde{\Theta}} - P_{1} P_{\widetilde{\Theta}}$
is defined similarly as $\sigma_{1,\Theta}$ in Eq. (\ref{sigma1Theta}).
The two types of forbidden configurations equivalent to that introduced for $\sigma_{1,\Theta}$ have form
\begin{equation}
F_{1,\widetilde{\Theta}}^{i_0} =
\frac{(\pi-2) (\pi^2 - 8)}{\pi^7 r^2}
+\frac{(\pi - 2) (16 - 24 \pi + 7 \pi^2)}{\pi^6 r^4}G_{0,0}
+ \frac{48 - 32 \pi - 30 \pi^2 + 26 \pi^3 - 5 \pi^4}{4 \pi^7 r^4} + \ldots
\end{equation}
and
\begin{equation}
F_{1,\widetilde{\Theta}}^{j_0} = - \frac{2(\pi-2)(4-\pi)}{\pi^6 r^4} \left(1 - 2\pi G_{0,0} + \log r + \gamma + \frac{3}{2} \log 2\right) + \ldots\,.
\end{equation}
In addition we also have to take into account correlations of defect $B_1$ with $T_1$, $T_2$, $T_3$, $T_4$, $T_5$, $T_6$ and $T_7$ in Fig. \ref{Tconfig}.
Their total contribution gives
\begin{equation}
\sigma_{1,T} =- \frac{6 (4 - \pi) (\pi-2)}{\pi^7 r^2}
+\frac{12 (4 - \pi) (\pi-2)}{\pi^6 r^4}G_{0,0}
+ \frac{36 - 81 \pi + 48 \pi^2 - 11 \pi^3 + \pi^4}{\pi^7 r^4} + \ldots\,.
\end{equation}

Next, using the procedure described in the next section, we find
\begin{eqnarray}
\sigma_{1,\widetilde{\Theta}} &\!\!=\!\!& \frac{(\pi-2)^2}{\pi ^6 r^2}
+\frac{(\pi -2) (3 \pi -4) G_{0,0}}{\pi ^5 r^4}
-\frac{(\pi -2) \left(22-28 \pi +7 \pi ^2\right)}{4 \pi ^6 r^4}\nonumber\\
&\!\!-\!\!&\frac{3 (4-\pi) (\pi -2) }{\pi^6 r^4}\left(\log r + \gamma + \frac{3}{2} \log 2\right) + \ldots\,,
\end{eqnarray}
and finally
\begin{eqnarray}
\label{AnswerP13}
\sigma_{1,3}(r) &=& \sigma_{1,2}(r) + 3 ( \sigma_{1,\widetilde{\Theta}}
- F_{1,\widetilde{\Theta}}^{i_0} - F_{1,\widetilde{\Theta}}^{j_0}) + \sigma_{1,T} =\\
 &-& \frac{(\pi-2)(8-\pi)}{\pi^6 r^4} \left(\log r + \gamma + \frac{3}{2} \log 2\right)
 +\frac{(\pi -2) \left(40 - 2\pi - \pi^2\right)}{2\, \pi^6 r^4} + \ldots\,.
\end{eqnarray}
which again confirms exactly (\ref{sigma1a}) and yields the same value for $B$ ! We emphasize that the value of $B$ computed here from $\sigma_{1,3}(r)$ is an independent calculation from that of $\sigma_{1,2}(r)$ and further strengthens the validity of the conformal formulas. 

By using the identity $\sum_{a=1}^4 \sigma_{1,a}=0$, we then readily find
\begin{equation}
\label{AnswerP14}
\sigma_{1,4}(r) \simeq
\frac{(\pi-2)(\pi+4)}{\pi^6 r^4}\left(\log r + \gamma + \frac{3}{2} \log 2\right)
+\frac{(\pi-2) \left(\pi^2 - 4 \pi - 16\right)}{2\, \pi^6 r^4}.
\end{equation}

Comparing with (\ref{sigma1a}) thus shows a full agreement between the dominant contributions of our exact results
on the lattice and the logarithmic CFT predictions.
Despite the very specific forms of $\Delta_{1\Theta}$ and $\Delta_{1\widetilde{\Theta}}$,
the correlation functions $P_{1,a}$, $a=2,3,4$ are the first example where
logarithmic corrections to pair correlations can be calculated explicitly.

\section{The ETI procedure for the calculation of $\sigma_{1,\Theta}(r)$ and $\sigma_{1,\widetilde{\Theta}}(r)$}


Consider the function $\sigma_{1,\Theta}(r)$ defined by (\ref{sigma1Theta}).
The matrices involved in the summation contain three types of Green functions.
First type does not depend on variables $k$, $l$ or $r$ and typically relates two sites a few lattice spacings apart.
For these we use their exact values.
Second type depends on $r$, but does not depend on $k$ or $l$.
Here we can replace the Green function by its asymptotic expansion in $r$.
Third type depends on $k$ and $l$.
In this case, as $k$ and $l$ take all values, small and large, we cannot use expansions and so we leave them in their integral form.
They have the general forms $G_{k+a,l+b}$ or $G_{k-r+a,l+b}$ with $a$ and $b$ taking values $0$, $\pm1$, $\pm2$ with the restriction $|a| + |b| \leq 2$.
We can reduce the number of such quantities by using Poisson equation, namely $\Delta G = I$.
It allows to express $G_{k,l}$, $G_{k-2,l}$, $G_{k+2,l}$, $G_{k,l-2}$, $G_{k,l+2}$ in terms of $G_{k+a,l+b}$\,,
and $G_{k-r,l}$, $G_{k-r-2,l}$, $G_{k-r+2,l}$, $G_{k-r,l-2}$, $G_{k-r,l+2}$ in terms of $G_{k-r+a,l+b}$ with $a, b = 0,\pm 1$, $(a,b) \neq (0,0)$.

After expanding the determinants, taking the sums over all $D$ and $D_\llcorner$ and after some manipulations we obtain, up to $1/r^5$ terms,
\begin{equation}
\sigma_{1,\Theta}(r) = K(r) + \sum_{(k,l)}\left(\frac{Y_1(r,k,l)}{r}
+ \frac{Y_2(r,k,l)}{r^2} +\frac{Y_3(r,k,l)}{r^3} +\frac{Y_4(r,k,l)}{r^4}\right)\ldots\, ,
\end{equation}
where $K(r)$ is explicitly given by
\begin{eqnarray}
K(r) &\!\!=\!\!&
\frac{(\pi-2)^2(8-\pi )}{4 \pi^5 r^4}G_{0,0}
-\frac{(\pi-2)^2(8-\pi)}{4 \pi ^6 r^2}
+\frac{(\pi-2) (40 - 43 \pi + 12 \pi^2)}{8 \pi^6 r^4}-\nonumber\\
\noalign{\smallskip}
&& -\:  \frac{(\pi-2)^2(32-\pi)}{8 \pi ^6 r^4}\left(\log r+\gamma+\frac{3}{2}\log 2\right) + \ldots
\end{eqnarray}
In the summation over $(k,l)$, we use the two symmetries $k\to r-k$ and $l \to -l$, in such a way that the functions $Y_i(r,k,l)$ can be expressed as linear combinations of the following (divergent) sums
\begin{equation}
R(r) \equiv \sum_{k=-\infty}^{+\infty}\sum_{l=-\infty}^{+\infty}G_{k - r + a_1,\, l + b_1} G_{k + a_2,\, l + b_2} G_{k + a_3,\, l + b_3}
\label{sumGGG}
\end{equation}
with parameters $a_i, b_i = 0, \pm 1$, $(a_i, b_i) \neq (0,0)$, $i=1,2,3$\,.
After substituting the integral representation of the Green function (\ref{Green2}), we get a triple integral over $\beta_1$, $\beta_2$ and $\beta_3$ corresponding to the three Green functions in the sum.
Performing the trivial sum over $l$ and the integral over $\beta_3$, we obtain
\begin{equation}
R(r) = \sum_{k=-\infty}^{+\infty} \int\!\!\!\!\int_{-\pi}^{\pi} \frac{{\rm d} \beta_1 {\rm d} \beta_2}{32\pi^2}
\frac{t_1^{|k + a_1 - r|}e^{{\rm i} \beta_1 b_1}}{\sqrt{y_1^2 - 1}}
\frac{t_2^{|k + a_2    |}e^{{\rm i} \beta_2 b_2}}{\sqrt{y_2^2 - 1}}
\frac{t_3^{|k + a_3    |}e^{{\rm i} \beta_3 b_3}}{\sqrt{y_3^2 - 1}},
\end{equation}
where we keep $\beta_3$ as a shorthand for $-\beta_1-\beta_2$. We recall here the definitions of $t_i = t(\beta_i)$ and $y_i = y(\beta_i)$, namely $t(\beta) = y(\beta) - \sqrt{y(\beta)^2-1}$ and $y(\beta) = 2 - \cos{\beta}$.
Since the parameters $a_i=0,\pm 1$ and $r \gg 1$, we can divide the sum over $k$ into three parts $(-\infty,-1]$, $[1, r-1]$, $[r+1, +\infty)$ and
separately consider the points $k=0$ and $k=r$. The summation gives
\begin{eqnarray}
S(a_1,a_2,a_3)&\!\!=\!\!&\sum_{k=-\infty}^{+\infty}t_1^{|k + a_1 - r|} t_2^{|k + a_2|} t_3^{|k + a_3|} \nonumber \\
&\!\!=\!\!&
t_1^r \left(t_1^{-a_1} t_2^{\left|a_2\right|} t_3^{\left|a_3\right|}
+\frac{t_1^{1-a_1}t_2^{1-a_2}t_3^{1-a_3}}{1-t_1 t_2 t_3}
+\frac{t_1^{ -a_1}t_2^{1+a_2}t_3^{1+a_3}}{t_1-t_2 t_3}\right)+\nonumber\\
&& + \; t_2^r t_3^r \left(t_1^{\left|a_1\right|}t_2^{a_2} t_3^{a_3}
+\frac{t_1^{1+a_1}t_2^{1+a_2}t_3^{1+a_3}}{1-t_1 t_2 t_3}
-\frac{t_1^{1-a_1}t_2^{a_2}t_3^{a_3}}{t_1-t_2t_3}
\right).
\label{Sa1a2a3}
\end{eqnarray}
By the change of variables $\beta_i \to -\beta_i$, the exponential $e^{{\rm i}\beta_1 b_1 + {\rm i}\beta_2 b_2 + {\rm i}\beta_3 b_3}$ can be replaced by
its real part $\cos(\beta_1 b_1 + \beta_2 b_2 + \beta_3 b_3) \equiv \cos(\beta_1 (b_1-b_3) + \beta_2 (b_2-b_3))$.
By the same symmetry $\beta_i \to -\beta_i$ , we may restrict the integration domain to $[0,\pi]\times[-\pi,\pi]$ and include a factor 2.
Since the expression for $R(r)$ is also symmetric under $(a_2,b_2) \leftrightarrow (a_3,b_3)$ , we can do a corresponding symmetrization,
\begin{equation}
R(r) =\frac{1}{32\pi^2} \int_{0}^{\pi} {\rm d} \beta_1 \int_{-\pi}^{\pi} {\rm d} \beta_2
\frac{S(a_1,a_2,a_3) C(b_1,b_2,b_3) + S(a_1,a_3,a_2)C(b_1,b_3,b_2)}{\sqrt{y_1^2 - 1}\sqrt{y_2^2 - 1}\sqrt{y_3^2 - 1}},
\label{Rint}
\end{equation}
with $C(b_1,b_2,b_3) = \cos(\beta_1 b_1 + \beta_2 b_2 + \beta_3 b_3)$. As mentioned above, $R(r)$ is divergent, but the proper linear combinations giving the functions $Y_i(r,k,l)$ have finite values.

To obtain the behaviour of $R(r)$ for large $r$, we first note that $0 < t(\beta) \leq 1$ for $-\pi\leq\beta\leq\pi$ and its maximum is at $\beta=0$.
For small values of $\beta$ it behaves
\begin{eqnarray}
t(\beta) = 1 - |\beta| + \frac{\beta^2}{2} + \ldots\,.
\end{eqnarray}
Moreover, we have the following asymptotic estimate of $t(\beta)^r$ for large $r$ (see \cite{jpr} for the derivation)
\begin{equation}
\left|t(\beta)^r - e^{-r|\beta|}\left(1 + \frac{r|\beta|^3}{12} - \frac{r|\beta|^5}{96}\right)\right| \leq \frac{162 \, e^{-6} }{r^4} + O(r^{-6}).
\label{tasympt}
\end{equation}

The Eqs. (\ref{Sa1a2a3}) and (\ref{Rint}) show that there are two types of integrals to be calculated,
\begin{equation}
I_1 = \int_{0}^{\pi} {\rm d} \beta_1 \int_{-\pi}^{\pi} {\rm d} \beta_2
\frac{t_1^r f_1(\beta_1,\beta_2)}{\sqrt{y_1^2 - 1}\sqrt{y_2^2 - 1}\sqrt{y_3^2 - 1}}
\end{equation}
and
\begin{equation}
I_2 = \int_{0}^{\pi} {\rm d} \beta_1 \int_{-\pi}^{\pi} {\rm d} \beta_2
\frac{t_2^r t_3^r f_2(\beta_1,\beta_2)}{\sqrt{y_1^2 - 1}\sqrt{y_2^2 - 1}\sqrt{y_3^2 - 1}}
\end{equation}
with regular functions $f_1(\beta_1,\beta_2)$ and $f_2(\beta_1,\beta_2)$.
From (\ref{tasympt}) we see that the main contribution to $I_1$ comes from the region of small $\beta_1$ and all $-\pi\leq\beta_2\leq\pi$.
For $I_2$, both $\beta_1$ and $\beta_2$ are small.
Next, we divide the integration over $\beta_2$ into three regions, namely,
\begin{eqnarray}
\mathbb{A} &\!\!=\!\!& [0,\pi]\times[0,\pi],\\
\mathbb{B} &\!\!=\!\!& \{(\beta_1,\beta_2): 0 \leq \beta_1 \leq \pi,\; -\beta_1 \leq \beta_2 \leq 0 \},\\
\mathbb{C} &\!\!=\!\!& \{(\beta_1,\beta_2): 0 \leq \beta_1 \leq \pi,\; -\pi \leq \beta_2 \leq -\beta_1 \}
\end{eqnarray}
and consider them separately.

\underline{\textbf{The region $\mathbb{A}$, $t_1^r$}}

Consider the series expansion for small $\beta_1>0$
\begin{equation}
\frac{f_1(\beta_1,\beta_2)}{\sqrt{y_1^2-1}} = \frac{c_{-1}(\beta_2)}{\beta_1} + c_0(\beta_2) + c_1(\beta_2) \beta_1 + c_2(\beta_2) \beta_1^2 + \ldots,
\label{f1series}
\end{equation}
where $c_i(\beta_2)$, $i=-1,0,1,2,\ldots$ are fixed regular functions.

Let us insert the expansion (\ref{f1series}) into the integral over $\beta_2$. It leads to a sum of terms of the form
\begin{equation}
Q(\beta_1)=\int_{0}^{\pi} {\rm d} \beta_2 \frac{c(\beta_2)}{\sqrt{y_2^2 - 1}\sqrt{y_3^2 - 1}},
\end{equation}
where $c(\beta_2)$ is one of the coefficients in the expansion (\ref{f1series}). This in turn can be written as
\begin{equation}
Q(\beta_1) =
\int_{0}^{\pi} {\rm d} \beta_2 \left[
\frac{c(\beta_2)-c(0)}{\sqrt{y_2^2 - 1}\sqrt{y_3^2 - 1}}
- \frac{\widetilde{c}(\beta_1)}{\sqrt{y_3^2-1}}\right]
+\int_{0}^{\pi} {\rm d} \beta_2 \left( \frac{c(0)}{\sqrt{y_2^2 - 1}\sqrt{y_3^2 - 1}} + \frac{\widetilde{c}(\beta_1)}{\sqrt{y_3^2-1}} \right)
\end{equation}
with
\begin{equation}
\widetilde{c}(\beta_1)= \left(
c'(0) - \frac{c''(0)}{2}\beta_1 + \frac{2 c'''(0)-c'(0)}{12}\beta_1^2+
\frac{c''(0)-c^{(4)}(0)}{24}\beta_1^3
\right).
\end{equation}

The virtue of this decomposition is that the first integral, with its substraction term, has a $\beta_1$ expansion with coefficients
which are regular functions of $\beta_2$ (up to order $3$ in $\beta_1$). It means that we can expand it over $\beta_1$ and then integrate
term by term over $\beta_2$, yielding a series over $\beta_1$. This trick was called Expand Then Integrate procedure (ETI, see \cite{jpr}).

The second integral of $Q(\beta_1)$ is divergent and yields a term proportional to the infinite constant $G_{0,0}$. To extract it, note that
\begin{equation}
\int_{0}^{\pi} \frac{{\rm d} \beta_2}{\sqrt{y_3^2 - 1}} =
\int_{\beta_1}^{\pi+\beta_1}\frac{{\rm d} \beta_2}{\sqrt{y_2^2 - 1}} =
\frac{3}{2} \log 2 - \log \beta_1 + \frac{\beta_1}{2\sqrt{2}} + \frac{\beta_1^2}{24} + \frac{\beta_1^3}{32\sqrt{2}}\ldots\,,
\end{equation}

\begin{eqnarray}
\int_{0}^{\pi}{\rm d} \beta_2
\left( \frac{1}{\sqrt{y_2^2 - 1}\sqrt{y_3^2 - 1}}
- \frac{1}{\sqrt{y_1^2 - 1}\sqrt{y_2^2 - 1}}
+ \frac{1}{\sqrt{y_1^2 - 1}\sqrt{y_3^2 - 1}} \right) =\nonumber\\
=-\frac{\pi -2}{4 \sqrt{2}} + \frac{3}{16}\beta_1 + \frac{(123 \pi -16)}{1536 \sqrt{2}}\beta_1^2 - \frac{11}{192}\beta_1^3+\ldots\,,
\end{eqnarray}
and
\begin{equation}
\int_{0}^{\pi} \frac{{\rm d} \beta_2}{\sqrt{y_2^2 - 1}} = 2\pi G_{0,0}.
\end{equation}
%
%
Therefore, we have, to order $\beta_1^3$,
\begin{eqnarray}
\int_{0}^{\pi} \frac{{\rm d} \beta_2}{\sqrt{y_2^2 - 1}\sqrt{y_3^2 - 1}}=
\left(
\frac{1}{\beta _1}
-\frac{\beta _1}{12}
+\frac{43 \beta _1^3}{1440}
\right) \left(2\pi G_{0,0} + \log\beta_1 - \frac{3 \log (2)}{2}\right)-\nonumber\\
-\frac{\pi }{4\sqrt{2}}+\frac{7 \beta _1}{48}+\frac{41\pi\beta_1^2}{512\sqrt{2}}-\frac{89 \beta_1^3}{1920}+\ldots\,.
\end{eqnarray}

\underline{\textbf{The region $\mathbb{A}$, $t_2^r t_3^r$}}

In this region $\beta_1>0$ and $\beta_2>0$. So we have $t_2^r t_3^r \sim e^{-\beta_2 r} e^{-(\beta_1+\beta_2)r} = e^{-(\beta_1+2\beta_2)r}$.
It means that we can replace  $t_2^r$ and $t_3^r$ by their asymptotics (\ref{tasympt}), make a double expansion of
\begin{equation}
\frac{f_2(\beta_1,\beta_2)}{\sqrt{y_1^2 - 1}\sqrt{y_2^2 - 1}\sqrt{y_3^2 - 1}}
\end{equation}
in $\beta_1$, $\beta_2$ and integrate them term by term.
%
%
After the integration over $\beta_1$ we obtain a series in $\beta_2$ starting with a $\beta_2^{-1}$ term. This term is singular, and yields $G_{0,0}$.
After carrying out the integral over $\beta_2$, we obtain a series in $1/r$, in which we keep terms up to $1/r^4$.

\underline{\textbf{The region $\mathbb{B}$}}

Here $-\beta_1 \leq \beta_2 \leq 0$, so $t_1^r \sim e^{-r\beta_1}$ and $t_2^r t_3^r \sim e^{r \beta_2}e^{-r(\beta_1+\beta_2)}=e^{-r\beta_1}$.
It means that we can expand both functions in the integrals $I_1$ and $I_2$ for small $\beta_1$.
After this, $\beta_2$ automatically becomes small and we can make a second series expansion.
To avoid the singularity in the denominator at $\beta_2=0$ and keep the variables in the region $\mathbb{B}$,
we should do first the expansion in $\beta_2$ around $\beta_2=-\beta_1$ and then the expansion in $\beta_1$ around $\beta_1=0$.
After the integration over $\beta_2$ from $-\beta_1$ to $0$, we come to a series in $\beta_1$ starting with a power $-1$.
Next we integrate term by term over $\beta_1$ and get a series by $1/r$, in which we keep terms up to $1/r^4$.

\underline{\textbf{The region $\mathbb{C}$}}

After the change of variables $\beta_2 \to \beta_3=-\beta_1-\beta_2$ the region $\mathbb{C}$ can be mapped into a subregion of $\mathbb{A}$.
In fact, we have
\begin{equation}
\int\!\!\!\!\int_{\mathbb{C}}{\rm d}\beta_1{\rm d}\beta_2 =
\int\!\!\!\!\int_{\mathbb{A}}{\rm d}\beta_1{\rm d}\beta_2 -
\int\!\!\!\!\int_{\mathbb{D}}{\rm d}\beta_1{\rm d}\beta_2,
\end{equation}
where $\mathbb{D}$ denotes the region
$\{(\beta_1,\beta_2): 0 \leq \beta_1 \leq \pi,\; \pi-\beta_1 \leq \beta_2 \leq \pi \}$. In this region
we can do a double expansion first in $\beta_2$ for $\beta_2 \approx \pi$ and then in $\beta_1$ for $0<\beta_1\ll \pi$.
Here we have no singularities and the integrations can be safely done term by term.

After the integration over $\beta_2$, we obtain a series in $\beta_1$ multiplied by $t_1^r$, which should be integrated term by term over $\beta_1$. Using the following integrals
\begin{eqnarray}
&& \int_0^\pi {\rm d}\beta_1 \, \frac{t_1^r}{\beta_1}           = 2\pi G_{r,0}    + {1 \over 12\, r^2} + \ldots \\
&& \int_0^\pi {\rm d}\beta_1 \, t_1^r\,\log{\beta_1}            =- \frac{\log{r} + \gamma}{r} - \frac{\log{r} + \gamma}{2\, r^3} + \frac{11}{12\, r^3} + \ldots\\
&& \int_0^\pi {\rm d}\beta_1 \, t_1^r                           = \frac{1}{r} + \frac{1}{2\, r^3} + \ldots\\
&& \int_0^\pi {\rm d}\beta_1 \, t_1^r\,\beta_1                  = \frac{1}{r^2} + \ldots \\
&& \int_0^\pi {\rm d}\beta_1 \, t_1^r\,\beta_1 \,\log{\beta_1}  = \frac{1}{r^2} \left(1 - \gamma - \log{r}\right) + \ldots\\
&& \int_0^\pi {\rm d}\beta_1 \, t_1^r\,\beta_1^2                = \frac{2}{r^3} + \ldots\\
&& \int_0^\pi {\rm d}\beta_1 \, t_1^r\,\beta_1^2\,\log{\beta_1} = \frac{1}{r^3} \left(3 - 2\,\gamma - 2\log{r}\right) + \ldots
\end{eqnarray}
we find the following explicit expansions,
\begin{eqnarray}
Y_1(r,k,l) &=& \frac{8-\pi}{\pi^2 r} - \frac{\pi+4}{2 \pi^2 r^3} + \ldots\,, \\
Y_2(r,k,l) &=& \frac{12\pi - 21}{\pi r^2} - \frac{24(\pi-2)}{\pi^2 r^2} \left(\log r + \gamma + \frac{3}{2}\log 2 \right) + \ldots\,, \\
Y_3(r,k,l) &=& \frac{8-\pi}{\pi^2 r} + \ldots\,, \\
Y_4(r,k,l) &=& \frac{2(8-\pi)(\pi-2)}{\pi^2} \left( -2\pi G_{0,0} + \gamma + \frac{3}{2}\log 2 + \log r \right) + \ldots,
\end{eqnarray}
leading to the announced result (\ref{sigma1ThetaResult}).

\section*{Discussion}

In our previous publication \cite{PhysLetterB} on the correlation function $\sigma_{1,2}(r)$ we have derived an expression for the leading asymptotics $A \log(r)/r^4$ with the coefficient $A$ coinciding with (\ref{Constant-A}). However, we later noticed two errors, compensating each other. The first error was the wrong sign at the sum in Eq. (\ref{N1Theta}).
The second one was omitting the contribution from the vicinity of $P_2$ in the sum (\ref{N1Theta}).
Surprisingly, these two errors exactly annihilate.
The correct and full derivation presented here needed much more elaborated calculations given in
Section VI and produced, as an essential by-product, the exact value of the coefficient $B$ in the expansion (\ref{sigma1a}). In addition, we extended the list of known correlation functions calculating $\sigma_{1,3}(r)$ and $\sigma_{1,4}(r)$.

While the two-point height correlations when at least one of the heights equals $1$ can be found using graph-theoretical methods, the correlations where both heights exceed 1 remain a difficult open problem. The method we used in this paper does not work anymore for topological reasons and therefore the extension of our calculations to these cases is highly non-trivial. At the same time, the correlations between $h_1$ and $h_2$ for arbitrary $h_1, h_2 = 1,2,3,4$ are necessary to establish a full correspondence between the LCFT prediction (\ref{pijgen}) and the lattice theory.

There are two other important problems, which remain unsolved. These are the problems of higher correlations and two-point correlations in the presence of a boundary, which should be most useful
to further check the correctness of the LCFT approach. Correlation functions of this kind with minimal heights have been obtained only on the square \cite{sand} and honeycomb lattice \cite{Ruelle-honeycomb}.

\section*{Acknowledgments}
This work was supported by a Russian RFBR grant No 06-01-00191a, and by the Belgian Interuniversity Attraction Poles Program P6/02,
through the network NOSY (Nonlinear systems, stochastic processes and statistical mechanics).
P.R. is a Research Associate of the Belgian National Fund for Scientific Research (FNRS).

\end{document}